**Spatially-resolved methane decomposition in a short glow discharge: insights into suprathermal hydrogen and radical chemistry**

Shurik Yatom

Princeton Plasma Physics Laboratory, Princeton University, Princeton, NJ 08543, United States of America

**Abstract**

This work presents the spatially resolved mapping of methane decomposition chemistry in a short direct-current glow discharge, revealing segregated zones for dissociation and polymerization in an Ar-$CH_4$ mixture at 400 mTorr. Laser-induced fluorescence (LIF), two-photon absorption LIF (TALIF), and optical emission spectroscopy (OES), are used to map the absolute number densities of atomic hydrogen (H), methylidyne (CH), and dicarbon ($C_2$) radicals. The results reveal a segregated chemical environment driven by the non-local electron kinetics. The primary dissociation of methane is confined to the cathode sheath (< 4 mm), where the H atom density peaks at a value (~$1.8\times10^{21}$ m$^{-3}$). Analysis of the $H_\alpha$ line profile identifies two distinct suprathermal H atom populations with peak kinetic energies of ~110 eV and ~17.5 eV, attributed to ion-surface reflection and electron-impact dissociation, respectively. The densities of secondary radicals CH and $C_2$ peak further from the cathode at the sheath-negative glow boundary (y≈ 4-6 mm), identifying this region as the primary zone for polymerization, with the peak $C_2$ density (~$5.6\times10^{17}$ m$^{-3}$) higher than that of CH (~$1.5\times10^{15}$ m$^{-3}$) by a factor of ~370. The discharge operates under strong non-equilibrium conditions, confirmed by the disparity between the low bulk gas temperature (~570 K) and the high vibrational temperature (3300-4700 K) of the emitting CH(A) radicals, the latter serving as a signature of a specific high-energy dissociative excitation pathway. This work demonstrates how the structured energy landscape of a short glow discharge spatially separates methane dissociation from subsequent polymerization. Kinetic estimates suggest that suprathermal H atoms may play a role in bulk plasma chemistry via abstraction reactions, with predicted $CH_4$ activation rates exceeding those of the thermal H population.

## 1. Introduction

Methane ($CH_4$), the main component of natural gas, is a key molecule in energy and chemical research due to its abundance and environmental impact. It is a vast hydrocarbon resource, making its efficient conversion into value-added products a paramount scientific and technological challenge[1]. Converting methane into hydrogen ($H_2$) and higher-value hydrocarbons (e.g., ethane, ethylene, acetylene) or carbon nanomaterials is a key research focus[2–7]. Traditional conversion methods, such as thermal pyrolysis or steam reforming, are energy-intensive[8], typically requiring elevated temperatures (often exceeding 1000 °C) and complex catalytic systems to overcome the high stability of the C-H bond.

Non-equilibrium, or “cold,” plasmas offer an alternative approach. In these weakly ionized gases, an external electric field energizes light electrons, creating a high-energy electron population while the bulk gas (ions and neutrals) remains near ambient temperature. This deviation from thermodynamic equilibrium enables electron-impact collisions to activate stable molecules like $CH_4$[9], bypassing high gas temperatures and enabling unique chemical pathways unavailable in thermal processes[10].

Glow discharge plasmas, a key type of non-equilibrium plasma, are used in processing[11], lasers[12], analytical chemistry[13,14], water treatment[15], gas conversion[16] and surface cleaning[17], including the surfaces on nuclear fusion reactors[18]. Methane-containing glow discharges are particularly relevant for depositing diamond-like carbon (DLC)[19–22]. This study focuses on a direct-current (DC) “short” glow discharge, where a small inter-electrode gap prevents the formation of a positive column (PC), creating a unique plasma environment with steep gradients in the electric field and electron energy.

The DC glow discharge, a self-sustaining plasma generated between two electrodes in a low-pressure gas, provides a convenient system for studying fundamental plasma phenomena. Its operation is predicated on a cascade of events initiated by the acceleration of free electrons, leading to ionizing collisions. The resulting positive ions are accelerated toward the cathode, where their bombardment induces the secondary emission of new electrons, which in turn sustain the discharge. A fully developed glow discharge exhibits a complex, stratified structure of luminous and dark regions along the inter-electrode axis. These regions include the cathode glow (CG), cathode dark space (CDS, also known as the Crookes or Hittorf dark space), the intensely bright negative glow (NG), the Faraday dark space (FDS), and, if the electrode gap is sufficiently long, a quasi-uniform PC.[23,24]

This work uses a configuration known as a "short" glow discharge, defined by an inter-electrode distance that is insufficient to support the formation of a PC[25,26]. The use of large, parallel-plate electrodes allows for the generation of a discharge that is laterally homogeneous and provides clear, unobstructed optical access for diagnostic techniques, a critical requirement for spatially resolved measurements. Finally, short discharges operating at low pressures (e.g., on the left-hand branch of the Paschen curve) typically exhibit a rising current-voltage (I-V) characteristic, which ensures stable, homogeneous operation that is free from the instabilities that can complicate experiments in other discharge regimes. Unlike PC or dielectric barrier discharges, where volume kinetics are often averaged, the short glow discharge regime creates a unique kinetic environment. It enables the spatial isolation of high-energy surface interactions (in the sheath) from volumetric ballistic transport (in the negative glow), allowing for the distinct identification of chemical drivers that are typically obscured in well-mixed reactor geometries.

This paper presents an experimental investigation into the spatially resolved chemistry of methane decomposition within a low-pressure (400 mTorr), large-area, short DC glow discharge sustained in an argon-methane mixture. This study employs optical emission spectroscopy to characterize the plasma. Laser spectroscopy, i.e., planar Laser-Induced Fluorescence (LIF) and Two-Photon Absorption LIF (TALIF), is employed to map the axial distributions of key reactive species: atomic hydrogen (H), methylidyne (CH), and dicarbon ($C_2$) radicals. The primary objective is to leverage the unique, structured environment of the short glow discharge to deconstruct the complex reaction network of methane.

The choice of these specific species is strategic for untangling the spatially dependent chemistry.

- **Atomic hydrogen (H):** As a direct product of the initial C-H bond breaking in methane, the spatial distribution of H atoms, measured via TALIF, serves as a direct probe of where the primary decomposition of $CH_4$ occurs. This allows us to distinguish between regions dominated by high-energy electron impact (expected in the cathode sheath) and those driven by energy transfer from argon metastables.

- **Methylidyne radical (CH):** This radical is a secondary or tertiary product of methane decomposition. Comparing its spatial profile, obtained via LIF, to that of atomic hydrogen provides crucial information about the subsequent steps in the chemical reaction chain and the transport of reactive species from their point of origin.
- **Dicarbon radical ($C_2$):** The $C_2$ radical is a key precursor to the formation of higher-order hydrocarbons and soot[27,27]. Its spatial distribution, measured by LIF, acts as an indicator for the onset of polymerization chemistry[28], revealing where in the discharge the pathways shift from simple fragmentation to C-C bond formation.

By correlating the distinct spatial profiles of these primary (H), intermediate (CH), and polymer-precursor ($C_2$) species with the non-local properties of the discharge, we aim to provide a fundamental understanding of how the spatially variant Electron Energy Distribution Function dictates the reaction pathways. Furthermore, the detailed spatial maps of these radicals will provide critical data to benchmark and validate kinetic models. The results will also guide future research by pinpointing regions of specific chemical activity.

## 2. Experimental setup and methods

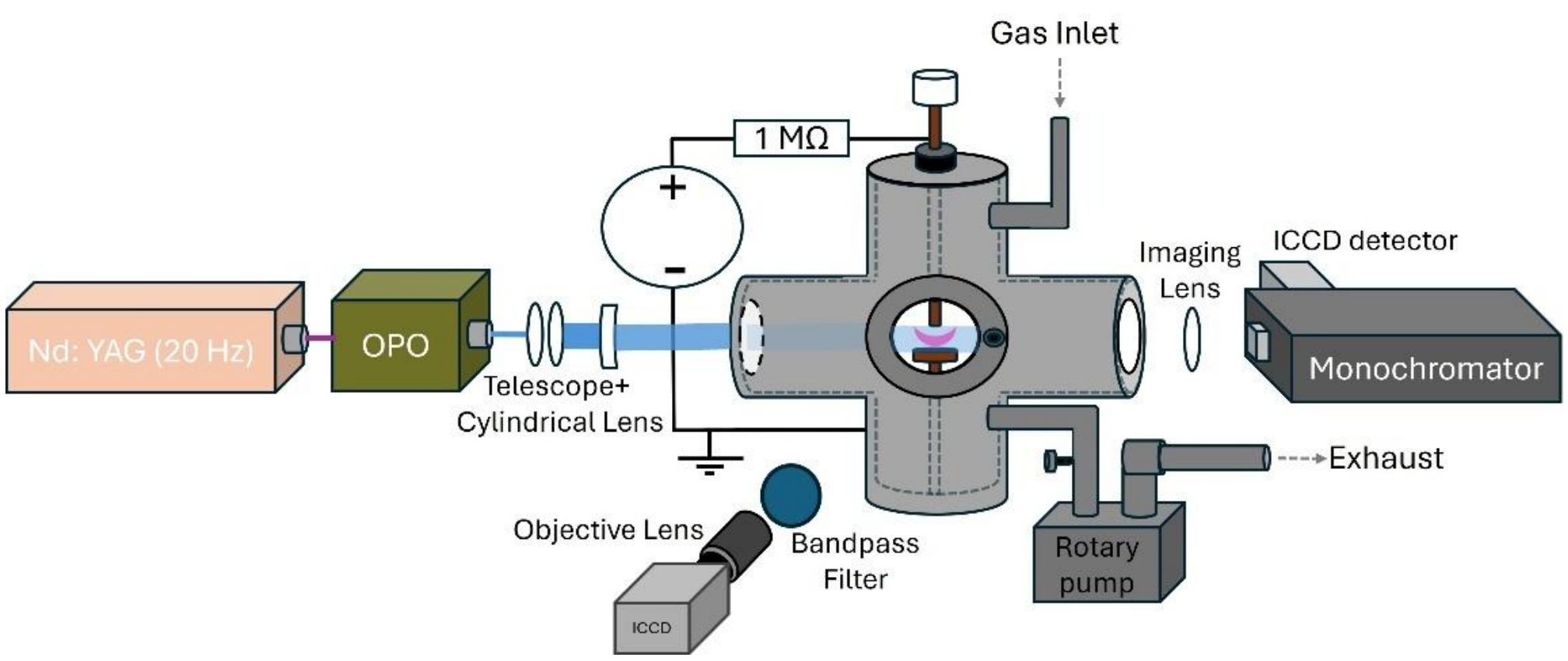

Figure 1. Experimental setup

The experiments were conducted in a stainless steel, 6-way cross chamber (8-inch diameter) with three fused silica windows for optical access and a fast-entry door with a window (Figure 1). The discharge was sustained between two parallel, coaxial electrodes: a 20 mm diameter copper anode, mounted on the top flange via a Wilson seal for vertical adjustment, and a 40 mm diameter brass cathode, fixed 17 mm below the anode and grounded. The anode, isolated by a quartz sleeve to prevent spurious discharges, was connected to a DC power supply (Glassman, 1 kV, 30 mA) through a 1 MΩ ballast resistor. The discharge current was 0.736 mA, calculated from the voltage drop across the resistor.

The chamber was evacuated to a base pressure of 30 mTorr using a rotary pump, then filled with a continuous flow of argon (Ar) and methane ($CH_4$) mixture (85% Ar, 15% $CH_4$) at 30 sccm and 5 sccm, respectively, regulated by mass flow controllers (Alicat Scientific). To minimize background gas contamination, the chamber was purged with the working gas mixture for at least 30 min. prior to data acquisition. During operation, the continuous flow of 35 sccm maintains a steady-state working pressure of 400 mTorr, continuously displacing residual species. The impact of the base pressure on discharge chemistry was assessed via optical emission spectroscopy: no emission lines attributable to $N_2$, NO, or OH were detected in the OES spectrum under discharge conditions. Furthermore, no $H_\alpha$ emission or atomic hydrogen signal was detected when operating the discharge in a pure Ar flow under otherwise identical conditions, confirming that any hydrogen-containing residual gas has a negligible contribution to the observed H atom production and chemistry. The working pressure was monitored by a Baratron capacitance manometer. The reactor's operation deposits thin carbonaceous films on the electrodes, requiring frequent cleaning with fine sandpaper and ethanol.

Two optical emission diagnostics were applied:

- **Filtered imaging**: An iCCD camera (PI-MAX 4, Princeton Instruments) with a 656.3 nm filter (1 nm FWHM) captured 2D $H_\alpha$ emission to estimate the discharge's axial current cross-section.
- **Spectral imaging**: The plasma was imaged onto a monochromator's entrance slit (HRS Spectra Pro 750, Princeton Instruments), coupled to an iCCD camera, producing spectra with axial spatial resolution[29]. CH(A-X) emission spectra were fitted using LIFBASE software to calculate rotational ($T_{rot}$) and vibrational ($T_{vib}$) temperatures. Calibration used a Hg-Ar lamp (Oriel 6035).

LIF and TALIF measured species densities:

- CH and $C_2$ LIF: A 20 Hz laser system (frequency-tripled Nd:YAG laser pumping an optical parametric oscillator at 355 nm) was used, as described in prior H-TALIF[30] and $C_2$-LIF studies[31].
- H-Atom TALIF: A 100 Hz diode-pumped laser (NT230, Ekspla) enabled faster signal collection for planar H-TALIF.

The laser beam was expanded (×3.75) using a two-lens telescope (f = 40 mm, f = 150 mm) and focused into a planar sheet (height: ~19.5 mm, waist: ~521 μm) using a cylindrical lens (f = 250 mm). Beam parameters and spectroscopic transitions are listed in Table 1. Fluorescence was imaged using an iCCD camera with a 3rd Gen HBf intensifier, an objective lens, and bandpass filters to isolate signals. Laser fluence (2.5-2.8 mJ/cm²) was kept low to avoid power broadening.

Table 1. Measured species and transitions.

| Species | $\lambda_{exc}$ (nm) | $\lambda_{fl}$ (nm) | Laser bandwidth (pm) | Beam energy (μJ) | References |
|---|---|---|---|---|---|
| H (1s→3s,3p,3d) | 205.13×2 | 656.3 | 12 | 250 | [32] |
| Kr($4p^6$→5p,3p,3d) | 204.2×2 | 587.1 | 12 | 250 | [33,34] |
| CH(A-X), $P_2(1)$ | 430.15 | 430 | 95 | 287 | [35] |
| $C_2$(d-a), $P_2(41)$ | 437.6 | 470 | 95 | 253 | [35] |

Fluorescence signals were imaged using an iCCD camera (PI-MAX 4, Princeton Instruments) with a Gen III HBf intensifier. The optical setup consisted of an objective lens and a bandpass filter, chosen to isolate the fluorescence signal of the measured species.

Our experimental approach is based on well-established methods from current literature. For Hydrogen Two-Photon Absorption Laser-Induced Fluorescence (H-TALIF) and its calibration using Krypton TALIF (Kr-TALIF), we employ a procedure similar to that of Niemi et al.[32] but using fluorescence at 587 nm rather than 826 nm, with the corresponding Einstein coefficient[36].Niemi's calibration framework remains applicable because the branching-ratio formalism depends only on the chosen decay channel's A coefficient, not on the specific transition wavelength. The 587 nm fluorescence is chosen due to the high quantum yield of the 3rd Gen HBf photocathode at visible wavelengths. This approach is routinely used for H-TALIF[30,33,34,37]. The diagnostics for $C_2$ and CH radicals using Laser-Induced Fluorescence (LIF) are based on the work of Luque et al[35]**.** To determine the rotational temperature in the interelectrode gap, we have used planar CH-LIF, pumping the rotational levels $R_2(1)$, $R_2(4)$, and $R_1(7)$ of CH(A-X) at respective wavelengths (426.93 nm, 428.66 nm, 430.15 nm) and observing fluorescence with a spectral filter (430 nm, 10 nm FWHM). These levels span a good range of J" levels and corresponding energies (~10.85-704.8 $cm^{-1}$), and their constants are well known and described in the literature. The background image was obtained by detuning the laser wavelength to 432.1 nm, where no significant absorption by CH is expected. The background contains the broadband emission from CH(A-X) and potential LIF signal from $C_3$, which is also excited in that wavelength range. The resulting corrected fluorescence signals were used for a Boltzmann plot, whose slope is used to determine the $T_{rot}$ value. The corrected fluorescence signals from pumping $R_2(1)$ and $R_1(7)$ levels were used as a two-line LIF temperature measurement to create a 2D gas temperature map, based on the assumption that $T_{rot}=T_{gas}$.

The fluorescence signal obtained by pumping CH(A-X) $R_2(1)$ level[38–42] was used to calculate the 2D absolute density of the CH radical. For density measurement of $C_2$ molecules, we have excited the $C_2$(d-a), $P_2(41)$ at 437.6 nm with partial overlap of $P_1(42)$ at 437.66 nm. The overlap is factored accordingly in the $f_B$ calculation. The fluorescence from (d–a)(2,1)(1,0) levels[31,35,43,44] is observed with a spectral filter (470 nm, 10 nm FWHM). The calibration of the instrumental parameters is done via Rayleigh scattering[45,46], where the laser wavelength was tuned to the fluorescent wavelength (430 and 470 nm, for CH and $C_2$, respectively). To correct for variations in detection efficiency, the spectral response of the iCCD camera and filter combinations was measured with a calibrated broadband light source (StellarNet SL1-Cal).

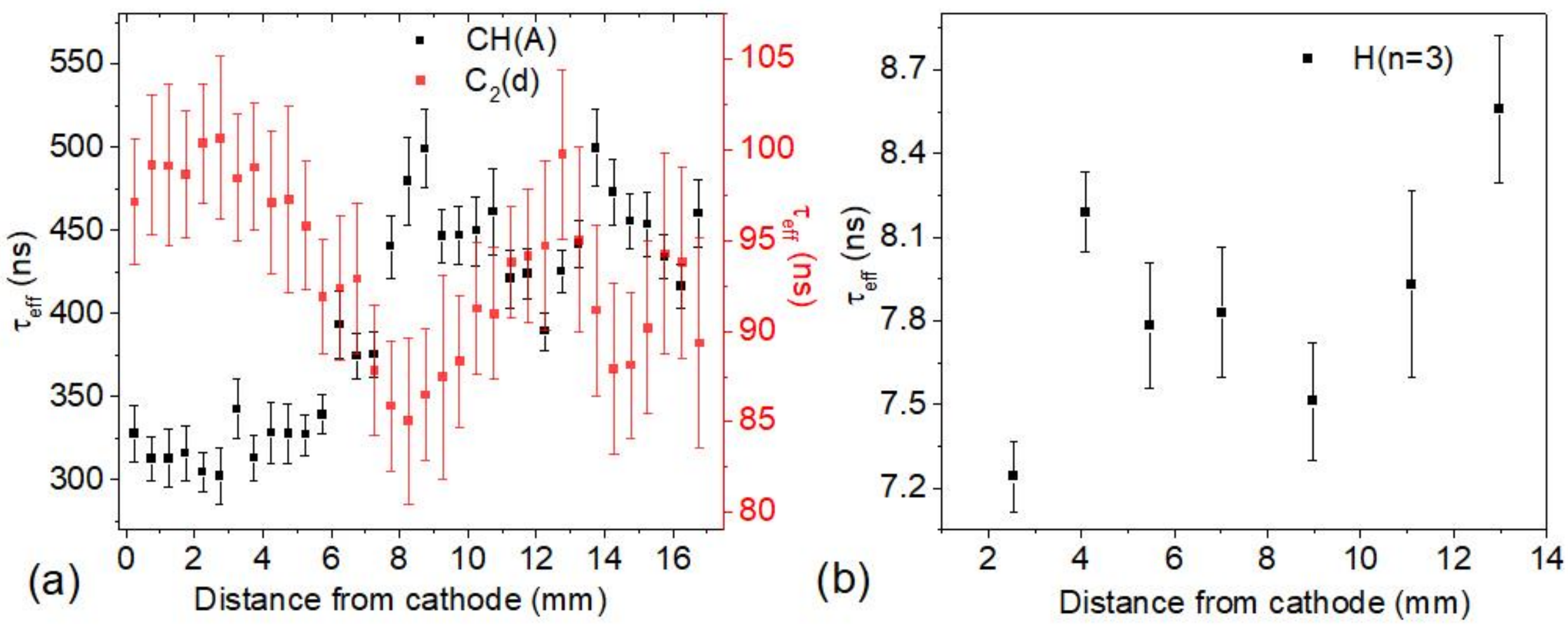

Figure 2. Effective TALIF lifetime vs height above cathode (a) CH(A) and $C_2$(d) (b) H(n=3).

Figure *2* presents the effective fluorescence lifetimes measured for $CH(A^2\Delta)$, $C_2(d^3\Pi_g)$, and H(n=3) across the discharge gap. The effective fluorescence lifetimes were measured for all species at different heights above the cathode using time-resolved fluorescence detection with short camera exposures and varying delays relative to the laser pulse (Figure *2*a). The ratio $\tau_0/\tau_{eff}$ defines the quantum yield ($\phi$), necessary for quantitative LIF interpretation. For CH(A) and $C_2$(d), planar LIF/TALIF provided spatially resolved measurements, while H-TALIF required conventional point measurements due to insufficient signal-to-noise ratio in planar mode. The axial dependencies of the effective lifetimes were integrated into the quantitative analysis of the planar LIF\TALIF images, assuming a radial uniformity. The molecular radicals exhibit opposing spatial behaviors, explained by three distinct quenching mechanisms. Measured $\tau_{eff}$ values span 531-350 ns for CH(A) and 102-76 ns for $C_2$(d), with upper limits approaching literature radiative lifetimes ($\tau_0$ = 535±10 ns for CH and 100-120 ns for $C_2$)[35,47,48].

To rationalize the opposing axial profiles- strong sheath quenching for CH(A) but minimal for $C_2$(d) - we propose three hypotheses grounded in charge-transfer reactions, symmetry considerations, and vibrational energy matching. These hypotheses may explain how species-specific interactions dictate the observed $\tau_{eff}$ variations.

Hypothesis 1: Ion quenching dominates CH(A) in the cathode sheath. $CH(A^2\Delta)$ shows strong quenching in the sheath (y < 4 mm) with $\tau_{eff} \approx$ 300-350 ns (quantum yields 57%-66%), attributed to charge-transfer reactions with energetic ions ($Ar^+$, $CH_x^+$). The heteronuclear CH radical's permanent dipole moment facilitates ion-molecule collisions with cross sections of 2-5×$10^{-20}$ $m^2$ at 200 eV[49,50]. Beyond y > 6 mm, $\tau_{eff}$ increases to 450-500 ns as ion density diminishes.

Hypothesis 2: $C_2(d^3\Pi_g)$ resists ion quenching due to symmetry and spin. $C_2$(d) shows minimal sheath quenching ($\tau_{eff} \approx$ 98-102 ns, quantum yields ~98%-100%) because: (1) no permanent dipole moment as a homonuclear molecule; (2) triplet-to-doublet charge transfer requires a spin flip with weak spin-orbit coupling in carbon[51]; (3) symmetric molecular orbitals reduce charge transfer efficiency. Ion quenching cross sections for $C_2$(d) are ~100× smaller than for CH(A).

Hypothesis 3: $CH_4$ dominates $C_2$ quenching via favorable energy matching. $C_2$(d) lifetime decreases to 85-90 ns in the y= 4-8 mm region, where $CH_4$ quenching dominates. The $C_2$(d→a) energy gap (1.2 eV, ~9,700 $cm^{-1}$) requires only 3-4 vibrational quanta of $CH_4$'s $\nu_3$ mode (3019 $cm^{-1}$), whereas CH(A→X) (3.1 eV) would require 8-9 quanta. Lower-order multi-quanta processes have exponentially higher probabilities. With $n_{CH_4} \approx 10^{21}$ $m^{-3}$ being $10^6$× higher than CH radical concentrations and $k_{vet}(CH_4) \approx 10^{-18}$ $m^3$ $s^{-1}$, $CH_4$ dominates the quenching rate despite CH's higher reactivity.

For H(n=3), measured lifetimes range from 7.2- 8.6 ns (quantum yields 70%-85%), lower than the theoretical 10.02 ns for coupled p=3 sublevels, indicating moderate quenching. The flat spatial profile and lack of dependence on $CH_4$ partial pressure suggest electron-impact deexcitation without strong spatial gradients. For atomic hydrogen, the measured H(n=3) lifetimes range from 7.2 to 8.6 ns across spatial positions from 2-13 mm, with no significant variation observed when changing the partial pressure of $CH_4$ from 5% to 40% in the Ar mixture (Figure *2*b). These values are lower than the theoretical average lifetime of 10.02 ns for statistically weighted p=3 sublevels (assuming full coupling of 3s, 3p, and 3d states), suggesting moderate quenching throughout the discharge with quantum yields of 70-85%. The relatively flat spatial profile indicates that the dominant quenching pathways for H(n=3)- likely electron-impact collisional deexcitation- do not exhibit the same strong spatial gradients observed for the molecular radicals.
Another important comment to make is the width of the absorption line, which is important for the determination of the overlap integral for both TALIF and LIF approaches. The laser bandwidth at TALIF wavelength (~204-205 nm) was measured in Kr gas at 375 mTorr – 12.3±0.53 pm. The absorption of H-TALIF is similar (Figure *3*). The limited tuning capability of the laser OPO system (10 pm) is a caveat, for it does not allow for precise measurement of small hot tails in the absorption profiles.

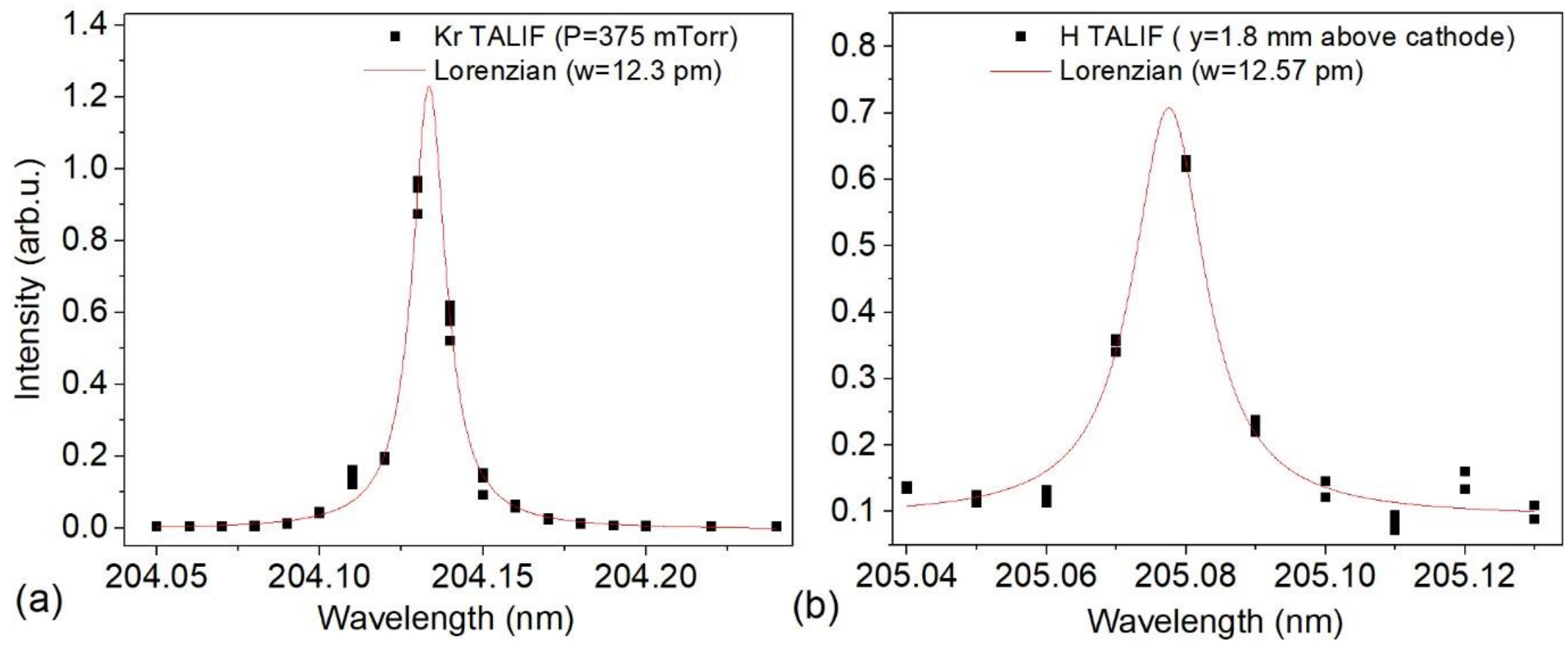

Figure 3. TALIF absorption scan (a) Kr, P=375 mTorr (b) H, y=1.8 mm above cathode, experimental conditions.

## 3. Results

The following section provides the results obtained by different diagnostic approaches, combined with a literature background that justifies those approaches. The detailed discussion of the results, with accompanying calculations and estimations, is contained in Section 4. Discussion.

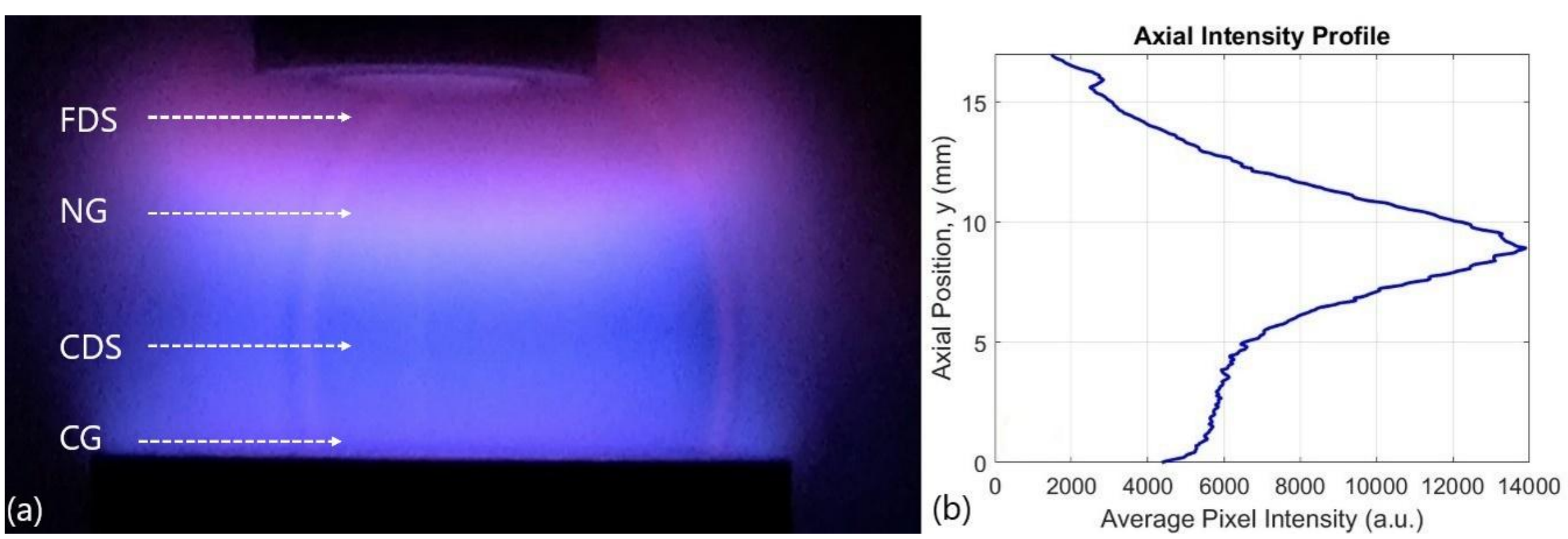

Figure 4. (a) Photo of the discharge with annotated discharge regions: Cathode Glow (CG), Cathode Dark Space (CDS), Negative Glow (NG), and Faraday Dark Space (FDS). (b) “White” light intensity distribution in the gap.

### 3.1. Imaging: conventional, filtered, and spectral.

A photo of the plasma in the interelectrode gap is presented in

Figure 4.a, with 2 distinguishable luminous areas: cathode glow (CG) adjacent to the cathode and a negative glow (NG), separated by the cathode dark space (CDS). The NG is curved, probably owing to the asymmetry of the anode and cathode dimensions, with the Faraday dark space (FDS) in between the NG and the anode.
Light intensity (

Figure 4.b) shows a prominent peak 9 mm from the cathode, corresponding to the center of the NG. Figure 5.a. shows an ICCD image of the discharge through a narrowband (1 nm FWHM) filter matching the H Balmer α line (656.3 nm). The region of $H_\alpha$ emission corresponds to the high $T_e$ and is associated with the plasma core, the hot, dense region where most of the discharge current is conducted. We use this image to determine the spatial cross-section along the axial direction and calculate the current density (see Figure 5.b). The volume occupied by plasma and the current density can be reliably estimated up to y=12 mm, but the main utility is for estimating the current density at the cathode and sheath boundary.

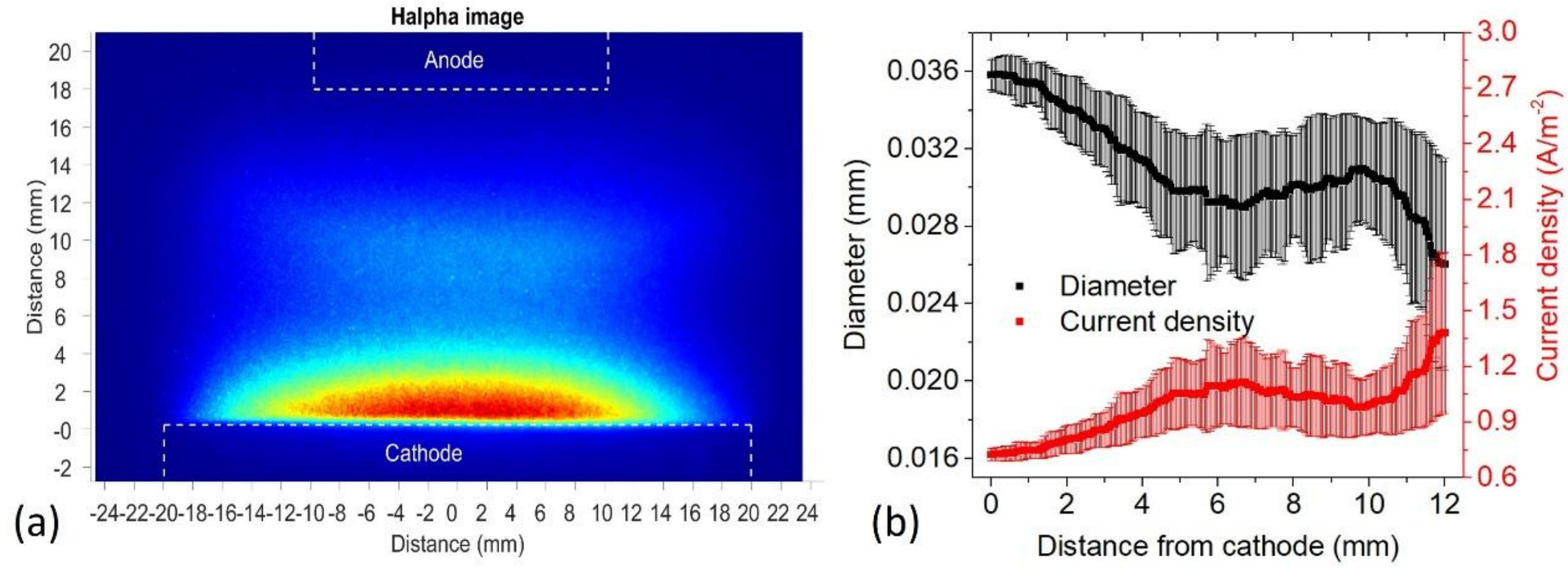

Figure 5. (a) Discharge imaged with a narrow filter (656.3 nm, 1 nm FWHM) (b) Discharge diameter and current density.

Spectral image of the $H_\alpha$ (Figure *6*.a) shows the strong broadening of the spectral line in the cathode region. Slicing the $H_\alpha$ image along the axial direction is useful for the detailed inspection of the line profile in the different regions. Here, we should note that the $H_\beta$ line was subject to the same analysis. The results of the inspection left us convinced that both $H_\alpha$ and $H_\beta$ are a composite of 3 Gaussian-like profiles with varying widths. Most of the analysis focused on the $H_\alpha$ line due its high signal-to-noise (SNR) ratio, contributing to confidence in fitting. Earlier works[52–55] reported that hydrogen Balmer lines (like $H_\alpha$ and $H_\beta$) observed in low-pressure gas discharges (typically in $H_2$, $D_2$ or Ar-$H_2$, Ne-$H_2$ mixtures) show "excessive" Doppler broadening, appearing as a complex profile with a narrow central peak and wide "wings". This broadening is caused by the presence of several distinct populations of hydrogen atoms, especially "hot" atoms with kinetic energies ranging from several eV to hundreds of eV. It is suggested that the fast atoms are primarily generated through a two-step process: first, hydrogen ions ($H^+$, $H_2^+$, $H_3^+$) are accelerated to high energies in the strong electric field of the cathode sheath; then, these energetic ions create fast neutral atoms either through gas-phase charge-exchange collisions with background molecules or, more significantly, through neutralization and backscattering from the cathode surface. Because the final line shape is a composite of these different populations (e.g., slow thermal atoms, atoms from dissociative excitation, and fast atoms from charge-exchange and backscattering), it is modeled by deconstructing the profile into a superposition of two or three Gaussian functions[52,56]. The characteristics of this broadening are dependent on the discharge conditions, the specific gas mixture, and especially the cathode material, which directly influences the efficiency of ion backscattering[54,56,57]. In this work, the cathode is made of Copper (Cu, Z=29). Studies that compare numerous cathode materials show that the intensity of the line wings follows a characteristic trend with the atomic number (Z) of the cathode. The material of the cathode in our experiments is Copper (Z=29), which, along with silver (Ag) and gold (Au), sits at a peak in this trend, producing stronger wings than many other materials[54].

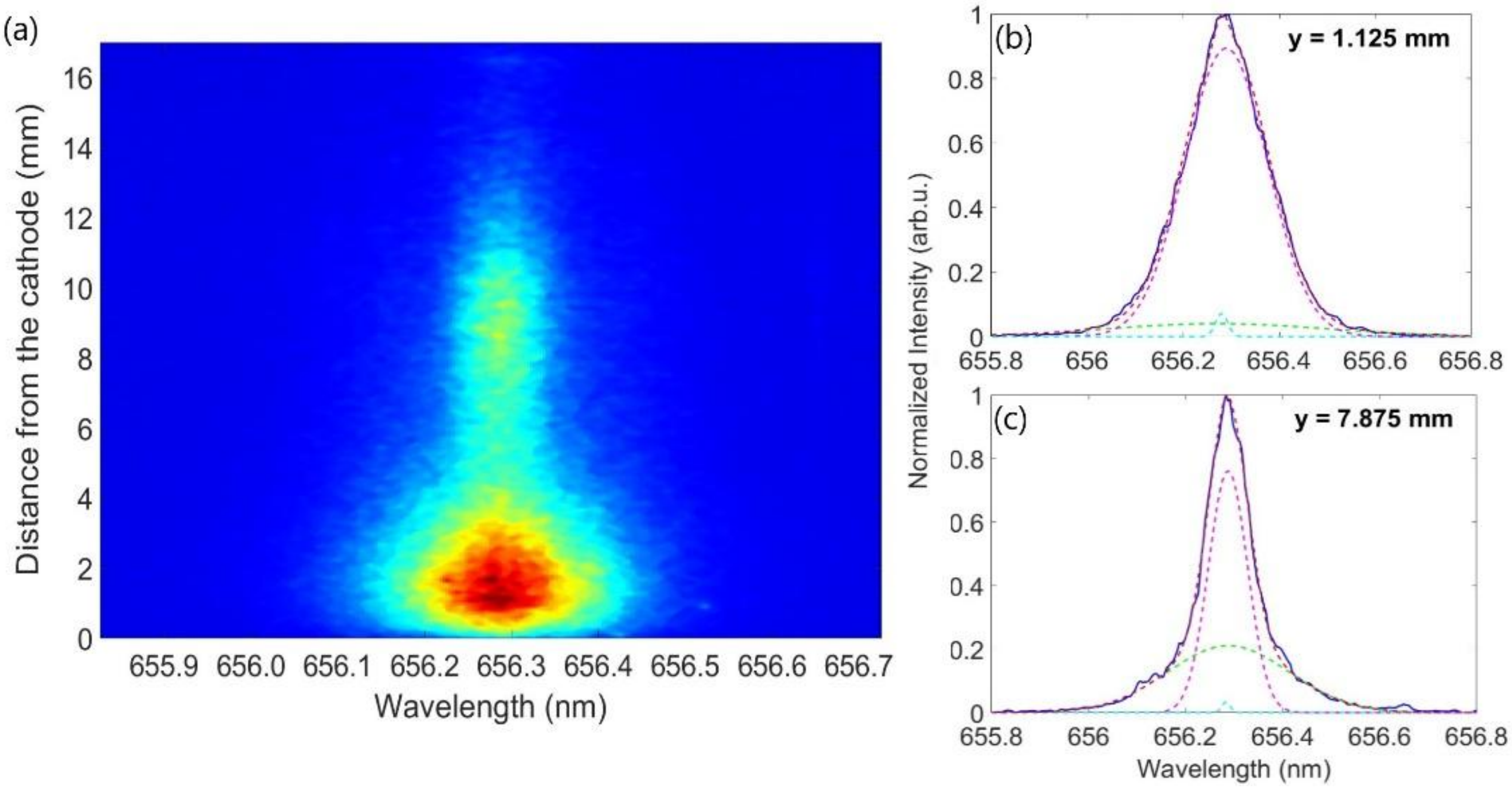

Figure 6. (a) Spatially resolved emission spectrum of the $H_\alpha$ line in the cathode-anode gap. (b) Fitting $H_\alpha$ spectra in different axial segments with 3 components.

In our case, the best fit for the $H_\alpha$ line shape is achieved with 3 peak functions: Broad Gaussian (BG), moderate Gaussian (MG), and a narrow Voigt (NV). Examples of modeled curves for two selected axial segments are shown in Figure *6*b and 6c. The iterative fitting procedure is described in Appendix 4.

The evolution of the amplitudes and FWHM for BG, MG, and NV is shown at Figure 7a and b, respectively. The results are compiled across 5 $H_\alpha$ spectral images corresponding to the same conditions, and the error bars reflect the statistical variation. The NV profile contributes to the triangular top and its width (~20-30 pm) changes little through the interelectrode gap, while its intensity shows a slow and steady decay from the cathode to the anode. The FWHM of NV corresponds to the Voigt width, from the convolution of the instrumental (Lorentzian ~ 15 pm) and "thermal" Doppler (Gaussian ~ 11 pm, Tg≈560

K, See Section 3.3). The widths and amplitudes of BG and NG show a dynamic evolution of two distinct populations of suprathermal hydrogen atoms to be discussed in Section 4.2.

It should be mentioned that Stark Polarization Spectroscopy method was applied to measure the electric field[58–60], but the π and σ polarized $H_\alpha$ and $H_\beta$ did not feature the peak separation of π-polarized line component, likely because the field is too low estimated as ~625 V/cm in the cathode sheath, see Section 4.1) to facilitate the SPS approach[58].

Another useful spectral image is captured in the range 423-437 nm (see Figure *8*.a) , featuring the emission of the CH(A-X) molecular system as well as several neutral and ion lines of Ar. Neutral Ar emission in that range originates from levels with energies of ~14.5-14.7 eV, while emission from ions emanates from levels with energies 19.3-21.35 eV. Inspection of the image and specific lines reveals that the emission peaks from ions and neutrals are spaced approximately 1 mm apart (Figure *8*.b).

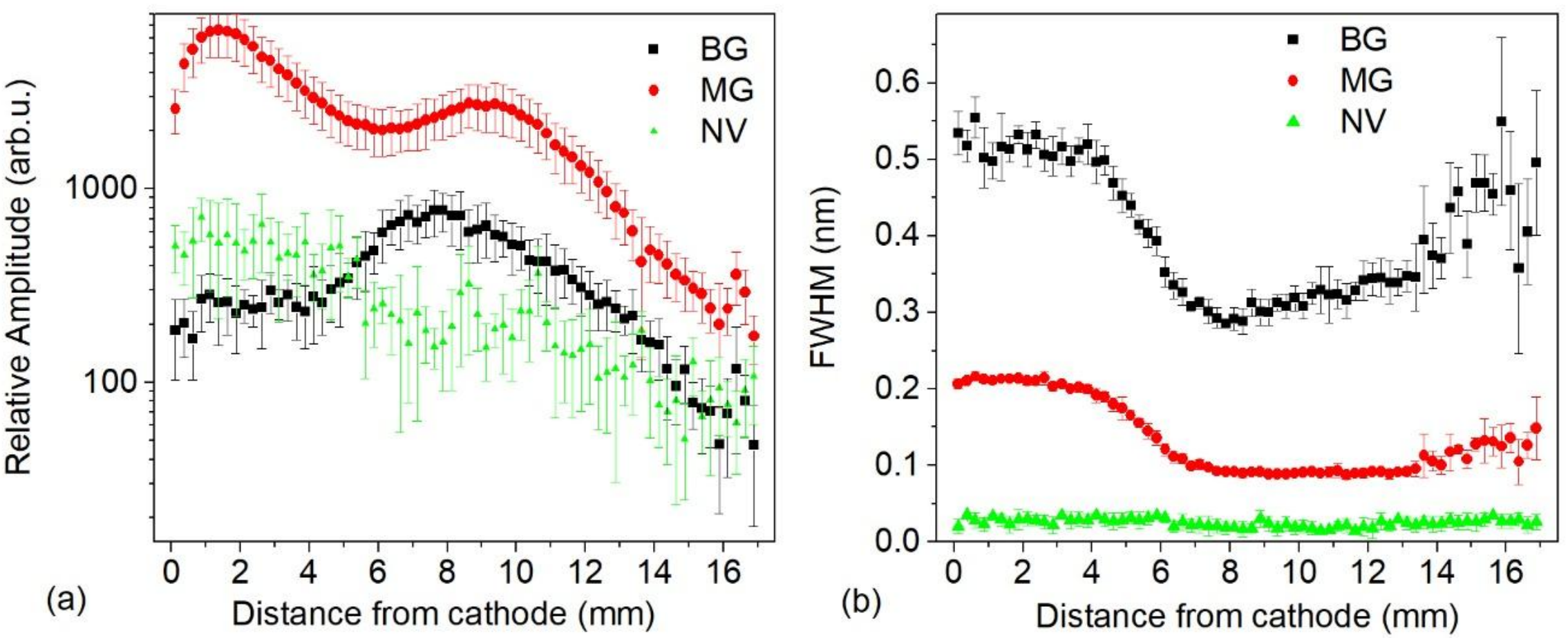

Figure 7 (a) Relative amplitudes and (b) FWHM of the 3-profile fit to $H_\alpha$ profile vs axial position.

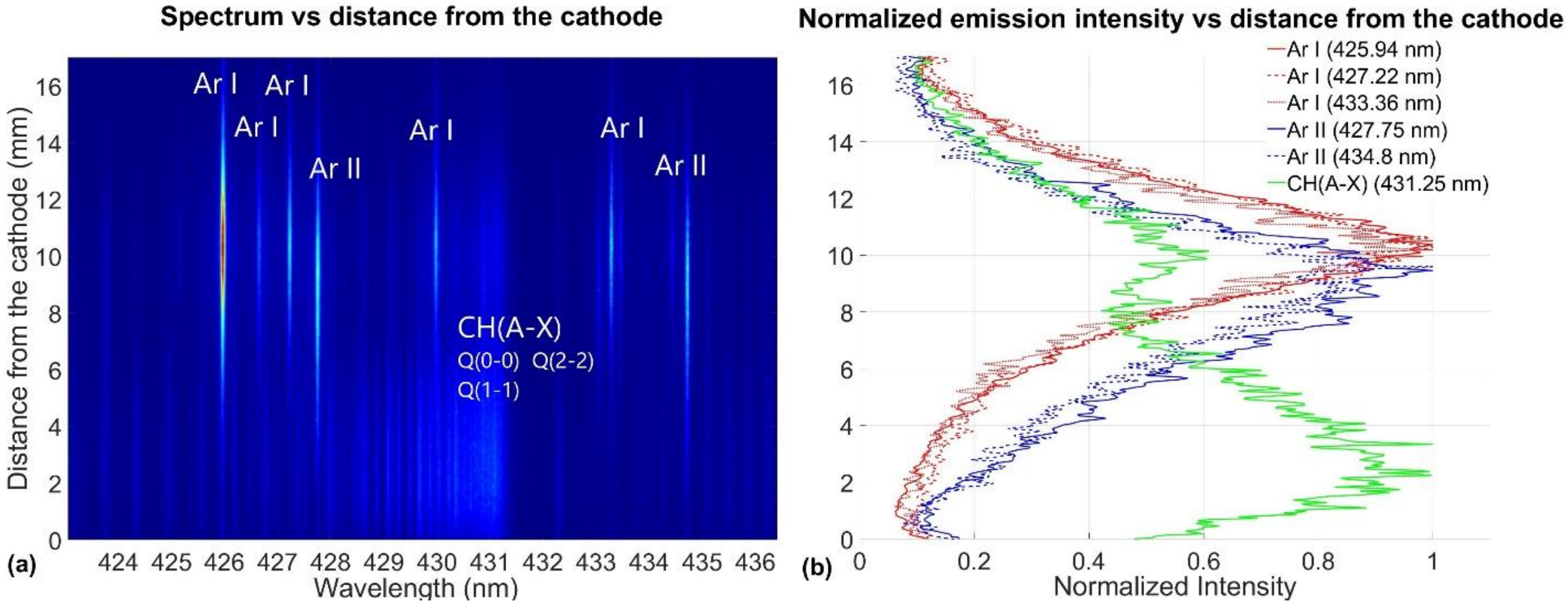

Figure 8. (a) Spectral image showing CH(A-X) band emission (Q (0-0) and Q (1,1) band heads at ~431.25 nm) and several lines from Ar neutrals (14.5-14.7 eV) and ions (19.3-22.6 eV). (b) Normalized emission intensity of CH(A-X) band head and selected Ar I and Ar II lines.

### 3.2. Nascent temperature measurements from CH(A-X) emission.

The emission of CH(A-X) in the spectral range of 426-436 nm features the Q(0-0), Q(1-1), and Q(2-2) band heads, as well as the R and P branches. Molecular emission is frequently used as a proxy thermometer, and in this case, the spectrum shows interesting evolution along the axial direction. The spectral image (Figure *8*.a) was sliced along the vertical axis, in intervals of 0.5 mm, and the resulting spectral curves were fit in LIFBASE software, see example at Figure *9*a. The fitting yields equivalent values of $T_{vib}$ and $T_{rot}$ at each segment, ranging from a high of 4700 K to a low of 3350 K. The vibrational temperature is mainly determined from the relative intensities of vibrational bands (e.g., $\Delta v = 0, 1, 2$) in the CH(A-X) system. Accordingly, the main procedure in our fitting algorithm is matching the intensity of the Q (2-2) band head by varying the $T_{vib}$ value and the "sloping" of the blue tail of the Q (0-0), Q ((1-1) bands from 4312.5 A to ~4300 A, in an iterative manner. This process results in a satisfactory matching of the spectrum, despite interference from the Ar I, II lines, especially further away from the cathode.

The evolution of the "emission" temperatures with distance from the cathode is shown in Figure *9*b. These measured values are non-characteristically high for bulk gas temperatures in glow discharges[61], likely stemming from a specific nature of formation of the excited CH(A) state[62–64]. The origin of the "hot" CH(A) state will be discussed in detail in Section **Error! Reference source not found.**.

### 3.3. Rotational temperature measurement by CH-LIF.

In contrast to emission from CH(A) level, CH-LIF thermometry probes the distribution at the ground CH(X) state, which is more relevant to probing the bulk gas parameters, $T_g$ in this case. Figure *10*.a shows an exemplary Boltzmann plot constructed from the spatially averaged LIF signals from a central segment in the plasma. The Boltzmann plot was generated by plotting the natural logarithm of the intensity divided by the degeneracy factor ($\ln(I/g_J)$) against the rotational energy ($E_J$) for each J'' level, where $g_J = 2J'' + 1$ accounts for the degeneracy, and $E_J \approx B_e \times J''(J'' + 1)$ with $B_e \approx 14.46$ cm⁻¹ for CH(X). The slope of the linear fit to these points is $-1/(k_B T_{rot})$, where $k_B$ is the Boltzmann constant (0.695 cm⁻¹/K) and the resulting temperature is $T_{rot}$=607±64 K.

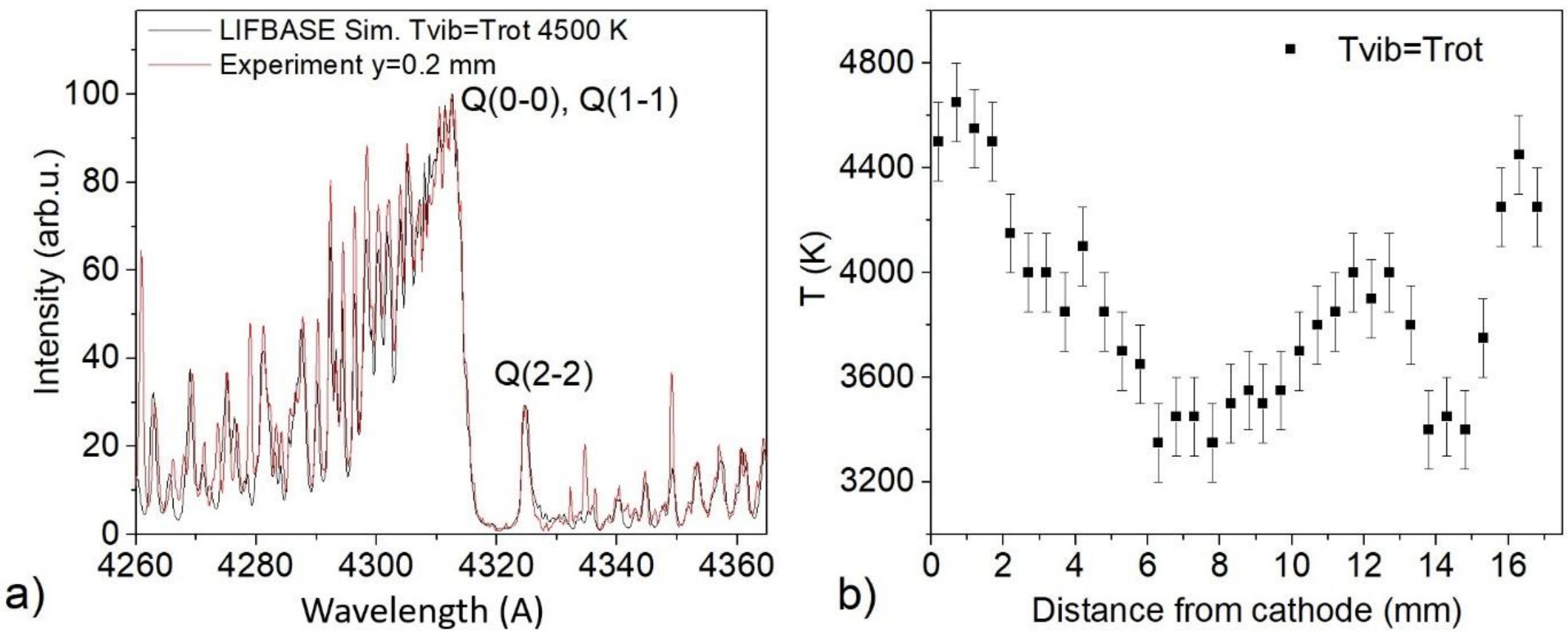

Figure 9. (a) Experimental spectrum at y=0.2 mm from the cathode and synthetic spectrum modeled with LIFBASE (b) $T_{vib}$ and $T_{rot}$ estimated from CH(A-X) emission versus the distance from the cathode.

For the 2D $T_{rot}$ map, we have used the signals from pumping $R_2(1)$ and $R_1(7)$ in a 2-line thermometry approach, owing to the ease of the computational approach for large matrices. The 2D $T_{rot}$ map shows a relatively uniform temperature profile, averaging to $T_{g,av}$=570 K across the discharge area. We consider this approach as a reliable measurement of $T_{gas}$ due to rotational equilibration in the ground state and probing of the ground state. With the estimated collision frequency $\nu \sim 2.27 \times 10^{6}$ s$^{-1}$ the characteristic time between collisions is $\tau_{mf}$~0.44 µs. We estimate that the residence time in the plasma is $\tau_{res}$~6.65 ms (see Appendix 1.B), ensuring that the CH(X) state equilibrates with the bulk gas. LIF probes this thermalized distribution, unlike the excited CH(A) state, where relaxation is limited by the short ~500 ns lifetime. LIF excites CH(X) to A and measures fluorescence, directly sampling the ground-state population without altering it significantly (low laser power avoids saturation). This contrasts with emission, which reflects the non-thermal CH(A) distribution from dissociative excitation (e.g., $e^- + CH_4 \rightarrow$ CH(A) + fragments), where $T_{rot}$(A) ~3300-4700 K is set by the excitation process[62–64], not $T_g$. Existing literature suggests that in similar situations, the ratio of "hot" to "cold" molecules is ~1/3[63–65]; hence, this ratio was adapted in this work to calculate the effective $T_{vib}$ and $T_{rot}$ values that are used for computing $f_B$ for $C_2$ and CH levels used in LIF diagnostics. The details and the calculations are given in Appendix 1.

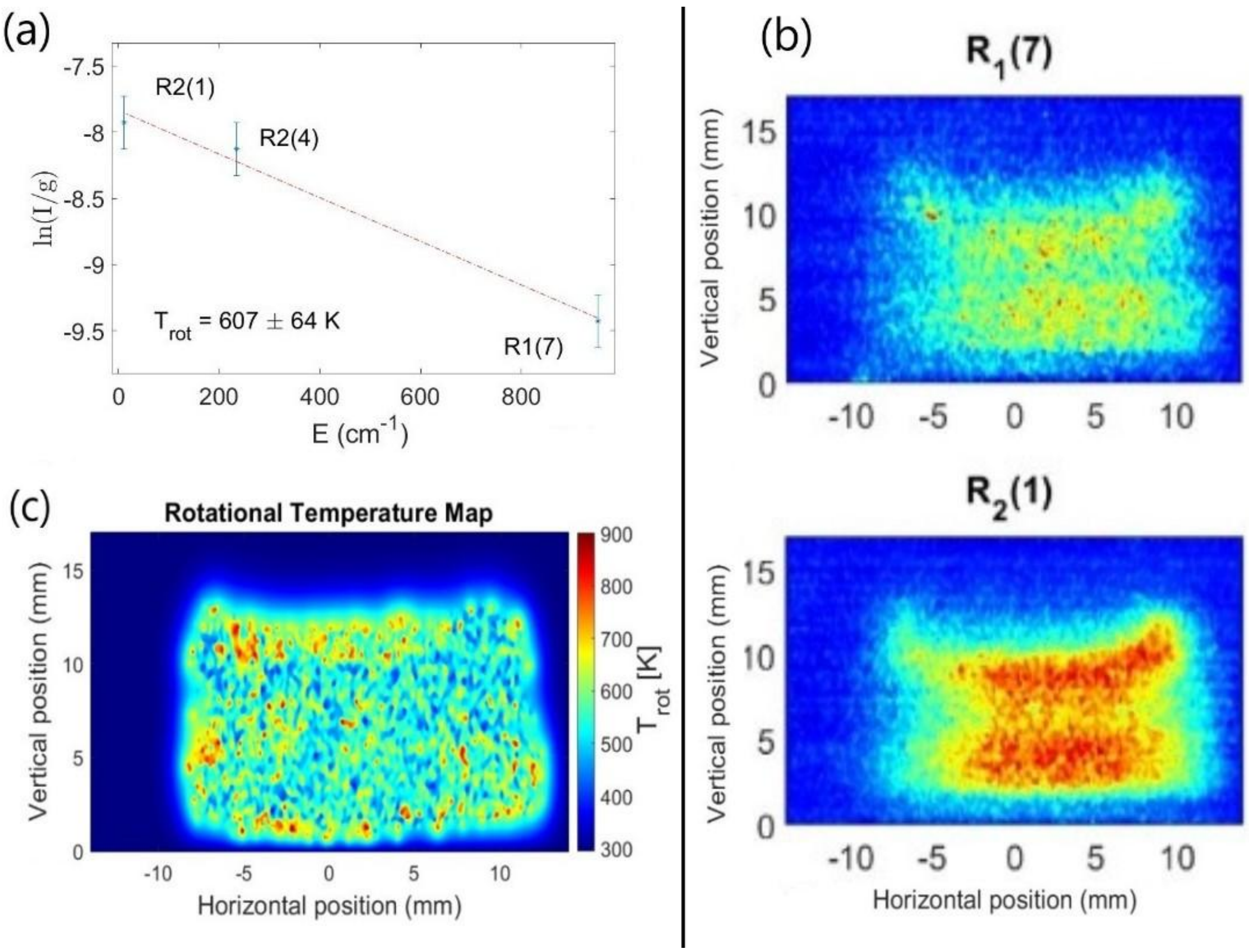

Figure 10. (a) An exemplary Boltzmann plot from pumping the $R_2(1)$, $R_2(4)$ and $R_1(7)$ levels (b) 2D LIF signal from $R_2(1)$ and $R_1(7)$ excitation (c) $T_{rot}$ map.

### 3.4. Density maps of CH, $C_2$, and H.

The 2D absolute number density maps for H, CH, and $C_2$ are shown in Figure *11*, with their corresponding axially averaged profiles (averaged from x = -6 mm to 6 mm) shown in Figure 12. The comparative analysis of these profiles reveals a segregated chemical structure within the discharge gap.

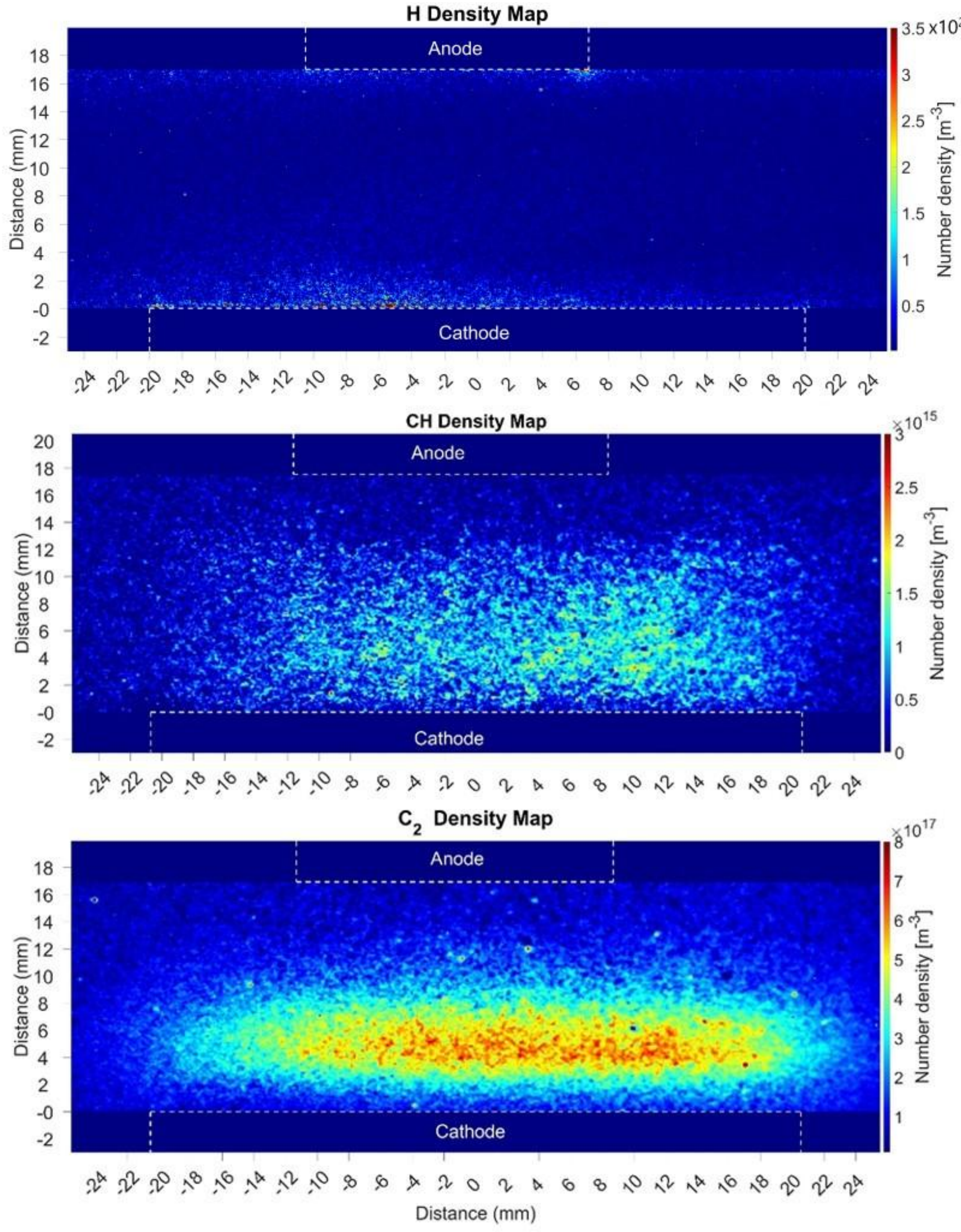

Figure 11. Density maps for H, CH, and $C_2$.

The atomic hydrogen (H) density is overwhelmingly confined to the immediate vicinity of the cathode. As shown in the 2D map (Figure *11*, top) and the axial profile (Figure 12, black line), the density exhibits a sharp, intense peak envelope of $1.8\times10^{21}$ m$^{-3}$ at y~0.2 mm, directly adjacent to the cathode surface. It is important to note that the density maps and profiles presented here are derived from raw, unsmoothed TALIF images. Consequently, the sharp, narrow spikes visible in the H density profile (e.g., at y~2.0 mm, and 16.8 mm) likely represent pixel-level intensity variations inherent to the intensified camera system rather than persistent, localized plasma structures. Following the near-cathode peak, the H density decays rapidly across the sheath region (y < 4 mm), dropping by nearly an order of magnitude before reaching the negative glow. A resurgence in H density is also observable towards the anode surface (y > 15 mm).

In stark contrast, the molecular radicals CH and $C_2$ are nearly absent in the high-density H region near the cathode. Their densities (Figure 12, red and blue lines) rise from the cathode, reaching their maximum values almost simultaneously at the sheath-negative glow boundary (y~4-5 mm). From this peak, both molecular radicals exhibit a gradual, congruent decay across the negative glow region toward the anode.

A notable finding is the significant disparity in absolute concentrations between these secondary radicals. The $C_2$ density (Figure 12, blue axis) reaches a maximum of ~$5.6\times10^{17}$ m$^{-3}$, which is more than two orders of magnitude greater than the peak CH density of ~$1.5\times10^{15}$ m$^{-3}$ (Figure 12, red axis), indicating an efficient chemical pathway for $C_2$ accumulation in this transitional zone. In the following "Discussion" section, we will examine this and other results in detail.

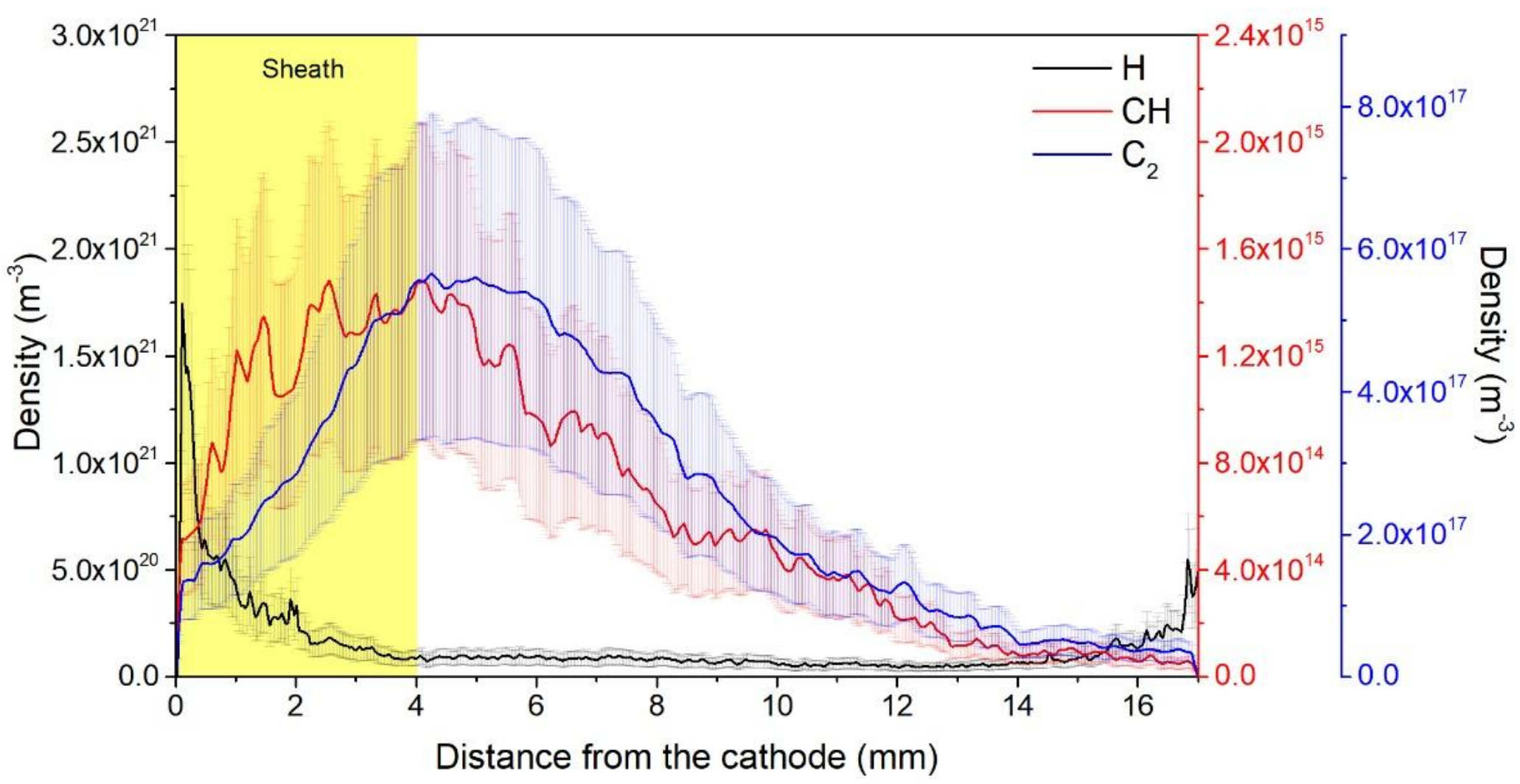

Figure 12. Averaged densities (-6 mm<x<6 mm) vs distance from cathode.

## 4. Discussion

### 4.1. Cathode sheath, collisionality, and plasma parameters

In this section, we will assess the collisionality of the cathode sheath to determine the regime of ion and electron transport, and to estimate the electron density ($n_e$) at the sheath edge and the cathode fall voltage ($V_c$). The sheath thickness $d_c\approx4$ mm, derived from $H_\alpha$ emission intensity profiles, serves as a key parameter. Collisionality is quantified by the ratio $d_c / \lambda_{mfp}$, where $\lambda_{mfp}$ is the mean free path (mfp) for electrons ($\lambda_e$) or ions ($\lambda_i$). This ratio indicates whether the sheath is collisionless ($d_c/\lambda\ll1$), collisional ($d_c/\lambda\gg1$), or transitional ($d_c/\lambda\approx1$-5).

The estimations are summarized in Table 2, which compiles all relevant cross-sections to be used in the following discussions. We have considered cross-sections for ions and electrons up to 100 eV and have identified $Ar^+$ and $CH_x^+$ as the dominant ions formed in Ar-$CH_4$ plasmas at low pressure. The results indicate a transitional sheath for ions ($Ar^+$, $CH_3^+$) and electrons (Elastic Scattering (~ few eV) is the relevant table section).

Table 2. Cross sections and mean free paths

| Process/Notes | Reaction | $\sigma(m^{-2})$ | $\lambda_{mfp}$ (mm) | $d_c/\lambda_{mfp}$ | Ref./Notes ($\sigma_{mix}=0.15\times\sigma_{CH4}+0.85\times\sigma_{Ar}$) |
|---|---|---|---|---|---|
| Charge exchange (CEX) (~200 eV) | $CH_3^+$ + Ar/$CH_4$ mix | $6.4\times10^{-21}$ | 23 | 0.17 | $CH_4$[49]<br>Ar[50] |
| | $CH_4^+$ +Ar/$CH_4$ mix | $4.9\times10^{-21}$ | 30.1 | 0.13 | |
| | $CH_5^+$+ Ar/$CH_4$ mix | $4.15\times10^{-21}$ | 35.54 | 0.11 | |
| Elastic Scattering (~ few eV) | $CH_3^+$ + Ar/$CH_4$ mix<br>$CH_4^+$ + Ar/$CH_4$ mix<br>$CH_5^+$ + Ar/$CH_4$ mix | $1.3\times10^{-19}$ | 1.13 | 3.5 | $CH_4$[66]<br>Ar[67] |
| | $Ar^+$ + Ar/$CH_4$ mix | $1.47\times10^{-19}$ | 1 | 4 | [68] |
| | $e^-$ + Ar/$CH_4$ mix | $5.22\times10^{-20}$ | 2.82 | 1.41 | [68,69,70] |
| Elastic Scattering for energetic H | H (100 eV) + Ar/$CH_4$ mix | $1.94\times10^{-19}$ | 0.76 | 5.3 | Ar[71] ×3 compensation for <20 mrad (forward glory), Polarization scaling for $CH_4$ |
| | H (17.5 eV) + Ar/$CH_4$ mix | $2.33\times10^{-19}$ | 0.63 | 6.36 | |
| Inelastic Scattering (~few eV) | $CH_3^+$ + Ar/$CH_4$ mix | $1.6\times10^{-20}$ | 9.21 | 0.43 | Ar[50] , $CH_4$[72,73] |
| | $CH_4^+$ + Ar/$CH_4$ mix<br>$CH_5^+$ + Ar/$CH_4$ mix | $3.2\times10^{-20}$ | 4.6 | 0.87 | |
| Electron impact dissociation (50-100 eV) | $e^-$+$CH_4$ | $0.8\times10^{-20}$ | 122.5 | 0.033 | [70] |

The cross sections in Table 2 are not sourced directly from experimental data for the exact reaction at the energy range (10–250 eV). They are stipulated or adjusted from analogous reactions using physical principles like IP mismatches, mass scaling, or energy dependence. The details are available in Appendix 2. The effective cross-sections were then derived from the respective cross-sections for $CH_4$ and Ar mixtures, using the formalism developed in Ref. [61].

We assume $T_e \sim 1.5$ eV at the cathode sheath edge. Such $T_e$ is a typical value for low-pressure glow discharges in noble gas-hydrocarbon mixtures. In pure Ar at similar pressures (0.1-1 Torr), electron temperatures range from 1-3 eV[61]. The addition of 15% $CH_4$ lowers Te slightly due to elevated inelastic cross-sections for dissociation and ionization (thresholds ~9-13 eV), which enhance electron energy loss through vibrational excitation and dissociative channels[74]. The low current density reduces heating, favoring a $T_e$ in the 1-2 eV range. This assumption aligns with experimental measurements in Ar-$H_2$/$CH_4$ plasmas at 0.1-0.5 Torr, where $T_e \sim 1$-2 eV at low power densities[59].

The electron density ($n_e$) is estimated at the sheath edge from the Bohm flux. The current density at the sheath's edge ($J_i$~1 A/m$^2$) is taken from *Figure 5*.b. The collisions in the sheath reduce the Bohm velocity[75] $v_B=\sqrt{\frac{k_B T_e}{m_i}}$ by a factor $\sqrt{1+\frac{\pi\cdot\alpha}{2}}$ , where $\alpha=\frac{\lambda_D}{\lambda_{mfp,i}}$, $\lambda_D=\sqrt{\frac{\varepsilon_0 k_B T_e}{n_e\cdot e^2}}$ is the Debye length. The algorithm here is to derive an initial $n_e$ from pure Bohm and use it for the estimation of $\lambda_D$ and α.

The $v_B$ is then corrected for the transitional regime and used to recalculate $n_e$ and re-iterate for $\lambda_D$ and α. The resulting values are $n_e \approx 2.39\times10^{15}$ m$^{-3}$, $\lambda_D$=0.184 mm and α=0.165.

The sheath is transitional for ions; therefore, we use the formalism from Lieberman and Lichtenberg[61], where the collisional cathode fall voltage for constant mobility $V_C$ is given by $V_C=\sqrt{\frac{8}{9}\frac{J_i d_c^3}{\varepsilon_0 \mu_i}}$. Under the high electric field conditions in the cathode sheath, ion mobility is significantly reduced from its low-field value. We utilize an effective high-field $\mu_i \approx 0.1$ m$^2$/V·s, typical for argon ions in this transitional regime[61], yielding a cathode fall voltage $V_c$=253 V which aligns well with the experimentally estimated plasma voltage $V_p \approx$ 264 V (from 1000V - $I_d \cdot R_{ballast}$ , with $I_d$=0.736 mA, $R_{ballast}$=1000 **Ω**), leaving ~ 11 V for the bulk plasma and anode sheath.

### 4.2. Suprathermal H

The lineshape of $H_\alpha$ shows an interesting dynamic of the 3 populations of hydrogen atoms described as BG (broad Gaussian), MG (moderate Gaussian), and NV (narrow Voigt). The intensities of the 3 components reflect the relative populations of atoms with corresponding kinetic energies. MG component accounts for the largest fraction of excited H atoms, manifested in their high intensity varying from ~7000 arb.u. in the sheath to ~200 in the vicinity of the anode. NV represents the excited population of the thermal H atoms. It marginally exceeds the intensity of BG in the sheath, but later subsides and becomes the lowest component, unlike the results of previous studies examining the abnormal Balmer H line broadening, where the thermal component is prolific and typically dominates[52,55,76,77]. It must be noted that the low intensity of NV does not mean that cold H atoms are not the dominant population. The H density measured by TALIF is essentially the cold thermal H atoms and the NV intensity is proportional to the excitation rate of H(n=1) to H(n=3) by electrons with energies 12-30 eV ( peak excitation cross-section)[72]. The MG intensity is due to the nascent, suprathermal H, created in the excited n=3 state, following the dissociation of $CH_4$ by e-impact. The observed result ($2<I_{MG}/I_{NV}<18$, Figure 7a) implies that the rate for the generation of nascent excited H exceeds the rate for excitation of thermal H to n=3 in the whole gap. Further quantitative analysis of this ratio requires knowing the density of $CH_4$ and the full EEDF throughout the gap. Lacking spatially resolved measurements or modeled values, this discussion is deferred to future work.

The NV maintains an almost constant width of ~20-30 pm throughout the entire interelectrode gap. Its width is likely defined by the Doppler broadening of thermal (~570 K, consistent with bulk gas temperature) atoms and instrumental function (~11 pm Gaussian and ~15 pm Lorentzian, respectively). The widths of MG and BG are much more substantial. Thermal Doppler and instrumental function account for only a tiny part of the measured widths. Stark broadening is negligible due to low electron density (~$10^{15}$ m$^{-3}$), as well as resonance broadening (due to low H density) and Van der Waals broadening (due to low pressure). Therefore, the width of the BG and MG after deconvolution of ~22 pm can be associated with Doppler Broadening from suprathermal H atoms. The kinetic energy (KE) associated with the observed widths and the intensity of the components are plotted in Figure *13*.a and b, for BG and MG, respectively.

The spatial evolution of these populations across the discharge reveals four distinct transport regimes, each governed by different physical mechanisms.

In the following analysis, the BG population is attributed to ion-surface reflection at the cathode and will be referred to as Reflected Suprathermal Hydrogen (ReSH). The MG population is attributed to electron-impact dissociation of $CH_4$ in the sheath volume and will be referred to as Electron-Impact Suprathermal Hydrogen (EISH).

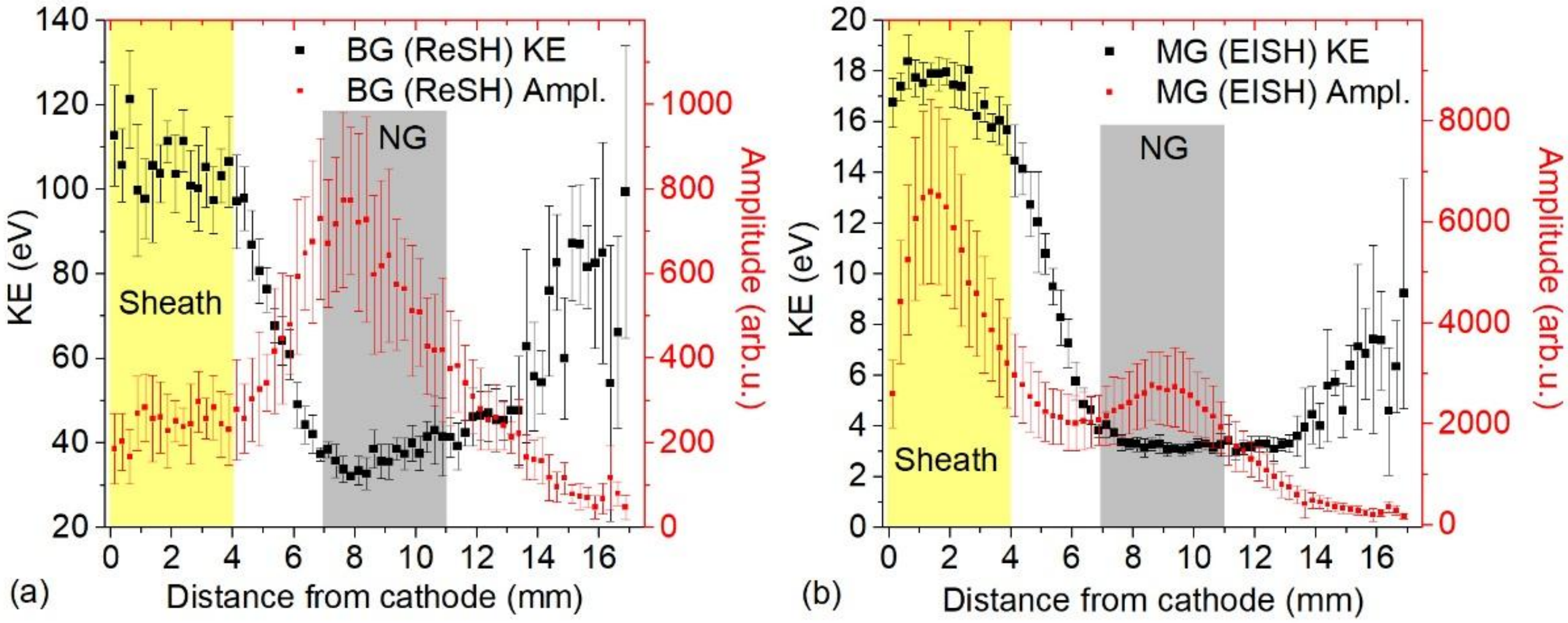

Figure 13. Kinetic energy of suprathermal H and relative intensity of BG (a) and MG (b) components vs distance from the cathode.

Important geometric note: the spectrometer measures Doppler broadening in the transverse direction, so only atoms with significant transverse velocity components ($\upsilon_{\perp}$) contribute to the observed signal. Atoms traveling predominantly along the discharge axis (the ballistic cohort) are effectively "dark" in this side-on geometry. Collisions that randomize trajectories convert this invisible ballistic population into the visible scattered population. The observed ⟨KE⟩ profile therefore represents a visibility-weighted measurement, not a simple energy average over all atoms.

#### 4.2.1. Zone 1: cathode sheath (0-4 mm)- production zone

The cathode sheath is the primary production zone for suprathermal populations. The sheath environment is defined by a cathode fall voltage of ~253 V over 4 mm thickness ($E/N \approx 9.4$ kTd), with $CH_4$ depleted by efficient dissociation, leaving a predominantly Ar atmosphere.

**RESH production and transport:** The ~110 eV population originates from ion-surface reflection at the cathode. Literature on Ar/$CH_4$ plasmas indicates $CH_5^+$, $CH_4^+$ and $CH_3^+$ as dominant ions due to ion-molecule reactions (e.g., $CH_4^+ + CH_4 \rightarrow CH_5^+ + CH_3$, rate ~$10^{-9}$ cm³ s)[61]. These hydrocarbon ions accelerated across the ~253 V cathode fall, impacting the copper surface where neutralization and dissociation occur (e.g., $CH_5^+ \rightarrow CH_4 + H$). Backscattered H atoms retain ~20-50% of the incident energy (reflection coefficient $R_n \approx 0.3$-$0.5$ for $CH_x^+$ on Cu[78]), yielding kinetic energies up to ~100-117 eV. The observed peak 110 eV reflects the transverse velocity component ($\upsilon_{\perp}$) in side-on geometry, averaged over the cosine-like angular distribution with a high-θ fraction (θ~60-90° from surface normal). The total KE per atom is thus higher (~145-220 eV in the distribution tail), consistent with effective energy reflection coefficients RE~0.4-0.6 for light H fragments from $CH_x^+$ dissociation and backscattering on rough Cu surfaces[53,79].The observation of this high-energy component in our transverse viewing geometry indicates the population is not a collimated beam; trajectory randomization occurs through diffuse scattering from the rough cathode surface and elastic scattering off heavy Ar targets, consistent with the "Collision Model" describing excessive Balmer line broadening in glow discharges[80] and validated by Monte Carlo simulations of heavy-particle transport[81].

The production rate is estimated at $\Gamma_{ref} \approx 0.45\text{-}8.4\times10^{19}$ $H\cdot m^{-3}s^{-1}$ (see Appendix 3), based on ion flux to cathode (~$4.6\times10^{15}$ $s^{-1}$), $CH_x^+$ fraction (50-70%), and reflection/fragmentation branching ratios. Charge exchange (CEX) is excluded as the primary source: effective mean free paths for CEX of energetic $CH_x^+$ (23-36 mm, see Table 2) exceed the sheath thickness, yielding <0.17 collisions.

Within the sheath, ReSH energy decreases only slightly from ~110 eV to ~100 eV over 4 mm. This stability reflects the $CH_4$ -depleted, Ar-rich environment where transport is dominated by elastic scattering. Consequently, the suprathermal atoms traverse this zone as a quasi-ballistic beam, with minimal energy loss or geometric removal. The relatively modest ReSH intensity observed in this zone partly reflects this directionality: atoms with strongly axial trajectories contribute minimally to the transverse Doppler signal.

This behavior is rooted in H-Ar collisions at high energies (>100 eV), which are predominantly elastic and forward-peaked due to the absence of internal energy sinks in atomic Ar, with differential cross-sections ($d\sigma/d\Omega$) favoring small deflection angles (average θ ~10-20°)[6,82]. The large mass mismatch ($m_H/m_{Ar} \approx 1/40$) limits fractional energy loss per collision ($\xi \approx 3\%$), preserving both axial and transverse velocity components ($\upsilon_\perp$) essential for side-on Doppler detection.

**EISH production and transport:** The ~17.5 eV population is produced by direct electron-impact dissociation of $CH_4$ in the sheath volume. Beam electrons (~50-100 eV) in the high E/N environment (~9 kTd) exceed dissociation thresholds (e.g., $e + CH_4 \rightarrow CH_3 + H + e$, ~8.8 eV; $e + CH_4 \rightarrow CH_2 + H + H + e$, ~12 eV; $e + CH_4 \rightarrow CH(A) + H_2 + H + e$; ~12.2 eV)[9,74,83–85]. These energetic electrons possess sufficient energy to access high-lying repulsive states of $CH_4$ (superexcited or Rydberg states) well above the dissociation threshold. Upon vertical Franck-Condon transition to a steep repulsive potential surface, the available energy is converted to fragment kinetic energy[9]. Conservation of momentum dictates that the light H atom acquires ~94% ($m_{CH3}/m_{CH4}$) of the kinetic energy release, while the heavy methyl fragment receives negligible recoil. The observed ~17.5 eV shelf is the characteristic signature of these specific high-energy repulsive pathways, consistent with experiments where H gains 10-20 eV from $CH_4/H_2$ dissociation[83,85,86].

The EISH atoms originate from the volumetric dissociation of methane, where fragments are likely ejected with angular anisotropy relative to the incident electron direction. This anisotropy is theoretically linked to the distortion of the initial $T_d$ symmetry toward $C_{3v}$ or $C_{2v}$ configurations on the dissociative potential surfaces[9]. Although such directional preferences are difficult to quantify and remain a recognized source of uncertainty in cross-section estimates[83], they are tentatively supported by the ground-state H-TALIF absorption scan at y = 1.8 mm (Figure *3*.b). The scan exhibits subtle intensity enhancements at approximately ±40 pm from the line center. This wavelength shift kinematically corresponds to the velocity boundaries of the ~17.5 eV EISH population. However, because these variations are near the noise floor of the measurement and are constrained by the OPO system's ~10 pm tuning limit and ~12.3 pm bandwidth, they cannot be definitively assigned to a suprathermal population without further uncertainty quantification. Therefore, while these variations offer an intriguing hint of a non-isotropic velocity distribution retaining a directional "memory" of the primary dissociation event, they currently remain a tentative observation. Definitive ground-state detection of this anisotropy will require future investigations utilizing higher-resolution, narrower-bandwidth laser systems. The potential significance of this observation is rooted in the transverse nature of the H-TALIF measurement. If verified, such features would specifically identify a cohort of fragments ejected with significant velocity components perpendicular to the discharge axis. These complex dynamics would otherwise be completely obscured in purely axial models.

The production rate estimated as $\Gamma_{e\text{-}dissoc} \approx 1.6\times10^{22}$ $H\cdot m^{-3}\cdot s^{-1}$ -approximately 100× greater than reflection (see Appendix 3), and CEX ($0.56\text{-}2\times10^{21}$ $H\cdot m^{-3}s^{-1}$, see Appendix 3) by ~10 due to high $n_e$ and $v_e$ (~$4.2–5.9\times10^{6}$ $m\cdot s^{-1}$). Literature confirms electron impact as the primary volume H source in $CH_4$ plasmas, with CEX contributing <10–20% to intermediate-KE populations[3,61]. The EISH energy remains nearly constant at ~17.5 eV across the sheath. As with ReSH, the $CH_4$-depleted environment and H/Ar mass mismatch preserve kinetic energy.

**4.2.2. Zone 2: transition region (4-8 mm) - rapid collisional energy dissipation.**

As suprathermal atoms exit the sheath, they enter the $CH_4$-rich transition region. Here, the strong electric field of the sheath decays to negligible levels, terminating the acceleration phase. Consequently, the transport regime shifts from field-driven ballistic flight to inertia-driven coasting, where the atoms' residual forward momentum is rapidly eroded by collisions.

Both populations undergo rapid energy loss: ReSH drops from ~100 eV to ~35 eV, while EISH decreases from ~15 eV to ~3-4 eV[87,88]. This massive energy dissipation marks the collision of fast H with methane molecules. This energy loss is driven by efficient inelastic energy transfer to $CH_4$ internal modes (vibrational/rotational excitation) and, at higher energies, to Collision-Induced Dissociation (CID) ($H_{fast}$ + $CH_4 \rightarrow H + CH_3 + H$).

This inelastic dominance in $CH_4$ enables the "crash" via efficient energy dissipation without strong trajectory randomization (forward-peaked scattering, average $\theta$~10-30$^0$ due to molecular recoil, preserving detectable $\upsilon_\perp$ components)[6]. Thus, while the transport begins to transition from ballistic to diffusive, the forward-peaked nature of the collisions limits geometric filtration losses, allowing the population to survive the energy crash before the massive removal in Zone 3[82].

Simultaneously, these inelastic collisions impart transverse velocity components, converting the spectroscopically "dark" ballistic population into a visible scattered population. The measured ⟨KE⟩ evolution in Zone 2 represents both (i) true energy loss from H-$CH_4$ inelastic collisions, and (ii) an apparent shift as the observation increasingly samples the cooled, scattered subpopulation rather than the still-energetic ballistic cohort.

The EISH intensity reaches a minimum at $y \approx 6$ mm, where atoms with ~3-4 eV efficiently drive $H+CH_4 \rightarrow H_2+CH_3$, creating a " chemical depletion zone" (abstraction rate ~$10^{-16}$ $m^3/s$ at 3-4 eV, see Section 4.4). The ReSH intensity continues to increase through this region, driven by a rising electron density (Ne) that enhances excitation efficiency, even as the absolute population decreases. EISH's lower KE makes them more susceptible to abstraction thresholds, explaining intensity dynamics.

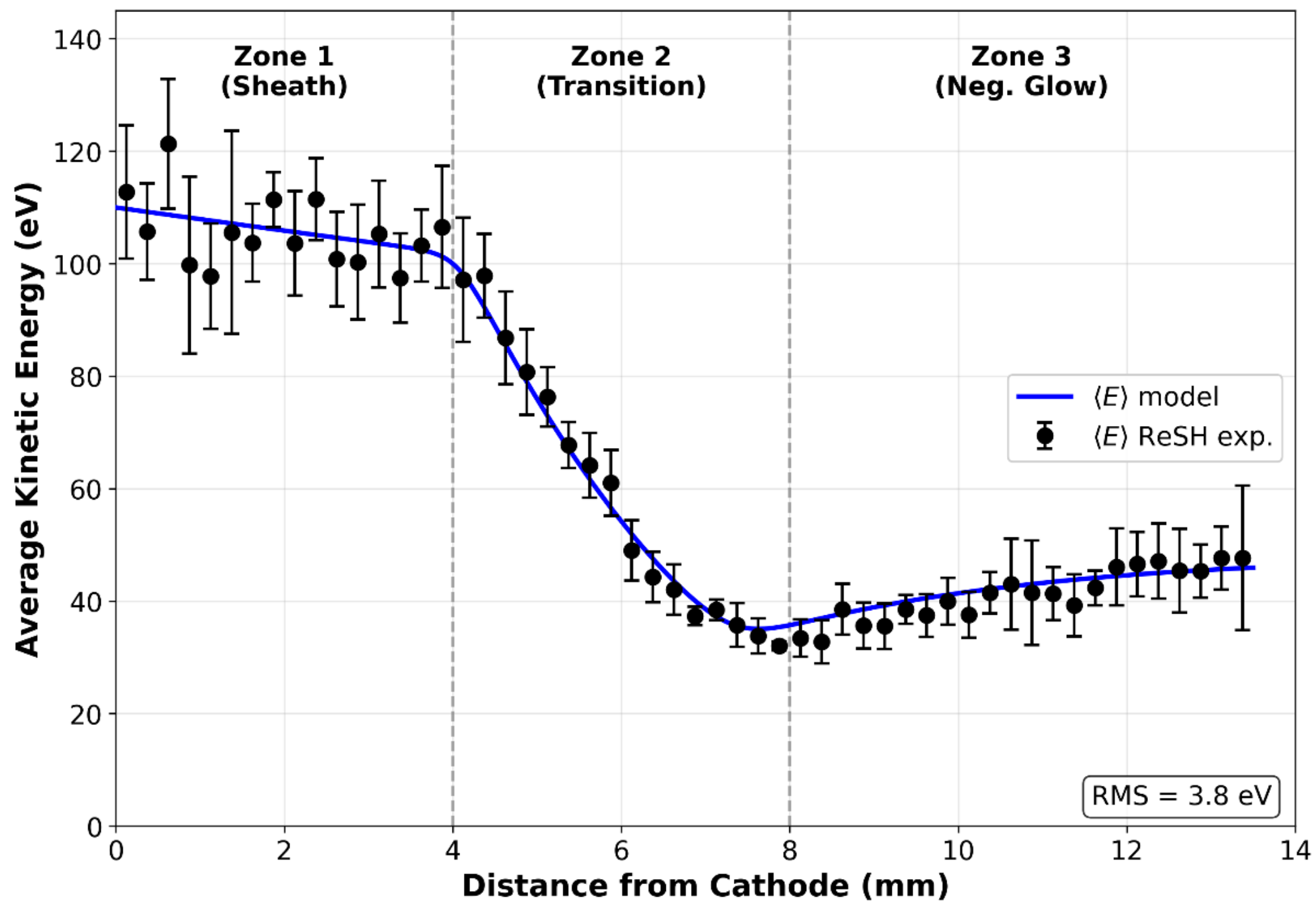

Figure 14. ReSH energy evolution: model (solid line) vs experiment (circles). The three transport regimes are evident: mild cooling in Zone 1 (sheath), rapid thermalization in Zone 2 (transition), and velocity filtration in Zone 3 (negative glow) where ⟨E⟩ rises from ~35 to ~48 eV despite ongoing collisions. RMS = 3.8 eV.

#### 4.2.3. Zone 3: negative glow (8-13.5 mm)- velocity filtration

In the negative glow, the kinetic energy dynamics of ReSH and EISH diverge dramatically. EISH remains constant at ~3.3 eV, continuously replenished by the same energetic electron population, which serves as evidence for a consistent tail temperature of EEDF. In contrast, ReSH kinetic energy rises from ~35 eV to ~48 eV while its amplitude drops. The abrupt change in ReSH energy evolution at y ≈ 8 mm- from rapid cooling to gradual rise- potentially signals a transition in the effective gas composition, possibly due to $CH_4$ depletion by electron density approaching its peak in the NG (~8-10 mm). Electrons, however, maintain sufficient $CH_4$ collision rates to sustain EISH production even at reduced $CH_4$ concentrations. EISH constancy highlights volumetric replenishment, less affected by filtration due to lower KE and isotropic birth[53]. This compositional transition is essential: without $CH_4$ depletion, the strong inelastic cooling ($\xi$ ~13-50% per H-$CH_4$ collision) would continue to dominate, overwhelming any filtration-driven KE rise. The observed transition from cooling to hardening at y ≈ 8 mm thus reflects a real change in gas composition, not merely a change in scattering dynamics.

The apparent re-heating of the ReSH is the signature of geometric velocity filtration. As the atoms cool, isotropic scattering preferentially removes the slower ones from the transverse observation volume. As demonstrated quantitatively by the 1D kinetic transport model detailed in Appendix 5, this energy-dependent geometric filtration artificially drives the average apparent energy of the surviving forward-peaked atoms upward (Figure *14*).

The ReSH emission peaks at y ≈ 8 mm, where rising $N_e$ maximizes excitation efficiency for the surviving ballistic population (forward-peaked at >60 eV)[82]. No new ReSH production occurs beyond the cathode. The EISH emission peaks later (y≈10 mm), reflecting renewed local production by energetic electrons (>13 eV) that persist deeper into the negative glow. Beyond y≈10 mm, both intensities decline monotonically as $n_e$ drops.

#### 4.2.4. Zone 4: FDS-anode region (y ≥ 13.5 mm)

Near the anode, both kinetic energy curves rise sharply: EISH increases to ~7.5 eV, and ReSH jumps to ~87 eV. This sharp jump is the "spectroscopic reveal" of a ballistic cohort that has traversed the gap nearly axially. Because the side-on spectrometer cannot detect purely axial velocity vectors, this cohort is "invisible" until it hits the anode, where diffuse reflection scatters atoms into transverse, "visible" trajectories. The observed ~87 eV represents the true energy this cohort possessed throughout its transit- minus modest losses from grazing H-Ar collisions (~10-15 eV over 17 mm, at ~3% per collision) and surface reflection (~5-10 eV). Quantitatively, extrapolating the Zone 3 trend predicts only ~55 eV at the anode- the observed ~87 eV exceeds this by ~30 eV, an excess that clearly indicates this distinct geometric reveal mechanism. The survival of this cohort at the anode strongly implies that $CH_4$ density in the gap is drastically reduced due to high dissociation rates near the cathode. In a $CH_4$-rich environment, inelastic collisions would rapidly destroy the ballistic population. These ~90 eV neutrals are energetically sufficient to drive sputtering of adsorbed hydrogen or carbonaceous species from the anode surface, typically considered passive in standard glow discharges. Emission intensities for both populations continue declining due to electron density depletion near the anode.

### 4.3. Spatially segregated chemistry and temperature

Figures 11 and Figure 12 show a discharge segregated into distinct chemical zones, driven by the non-uniform electron kinetics of a short glow discharge. Chemistry proceeds as a sequence of zone-specific processes: dissociation near the cathode, transitioning to polymerization at the edge of the negative glow. This non-equilibrium structure is further confirmed by the spatial profile of the emitting CH(A) vibrational temperature (Figure 9b). The profile exhibits elevated temperatures near both electrodes (peaking at ~4700 K near the cathode and rising similarly near the anode), with a broad minimum (~3300 K) in the intermediate negative glow region. These values are non-characteristically high for bulk gas temperatures ($T_g \approx 570$ K), serving as a direct fingerprint of the high-energy processes driving the local chemistry.

**Zone 1: the cathode sheath (y < 4 mm). The decomposition zone.**

In this zone, fast electrons accelerated across the large potential drop (estimated at ~250 V) possess energies in the tens of eV, enabling dissociative excitation of $CH_4$ via reactions such as $e^- + CH_4 \rightarrow CH(A^2\Delta) + H + H_2 + e^-$ (~12.2 eV)[83–85], which accounts for ~5-10% of total dissociation branching at 50-100 eV ($e^-$ energies corresponding to peak cross section for $e^-$ impact dissociation of methane). The branching ratios for electron-impact dissociation of methane at these energies favor pathways like $e^- + CH_4 \rightarrow CH_3 + H + e^-$ (~40–50%, threshold ~8.8 eV) and $e^- + CH_4 \rightarrow CH_2 + H + H + e^-$ (~20–30%, threshold ~12 eV), but the excited CH(A) channel is minor yet key for $T_{vib}$, as the excess energy above the threshold is preferentially partitioned into vibrational modes of the CH(A) fragment, resulting in non-thermal, elevated $T_{vib}$ values that exceed the gas temperature ($T_g \approx 570$ K) from CH-LIF ground-state measurements. The measured peak density of $\sim 1.8\times10^{21}$ $m^{-3}$ represents approximately 44% of the theoretical maximum H yield from complete $CH_4$ dissociation ($n_{Hmax} \approx 4.1\times10^{21}$ $m^{-3}$, calculated from the initial methane feedstock at 570 K). This high atomic fraction is consistent with efficient electron-impact dissociation in the sheath, supplemented by surface-driven hydrogen recycling from the ion-bombarded cathode. As ions continuously implant hydrogen into the copper lattice and the surface reaches saturation, the resulting desorption flux acts as a secondary hydrogen source that sustains the near-cathode density. The axial profile of thermal atomic hydrogen density $n_H(y)$ is well described by a double-exponential decay with constants $t_1 \approx 0.11$ mm and $t_2 \approx 1.1$ mm, see Figure *15*. The possible interpretation is of a narrow production layer (characteristic width $t_1 \approx 0.11$ mm) immediately adjacent to the cathode, where the high electric field accelerates electrons into the 50-150 eV range optimal for $CH_4$ dissociation, while ion neutralization and reflection simultaneously release additional H via fragmentation of excited $CH_x^{o*}$ neutrals on a comparable ~0.01-0.1 mm scale.

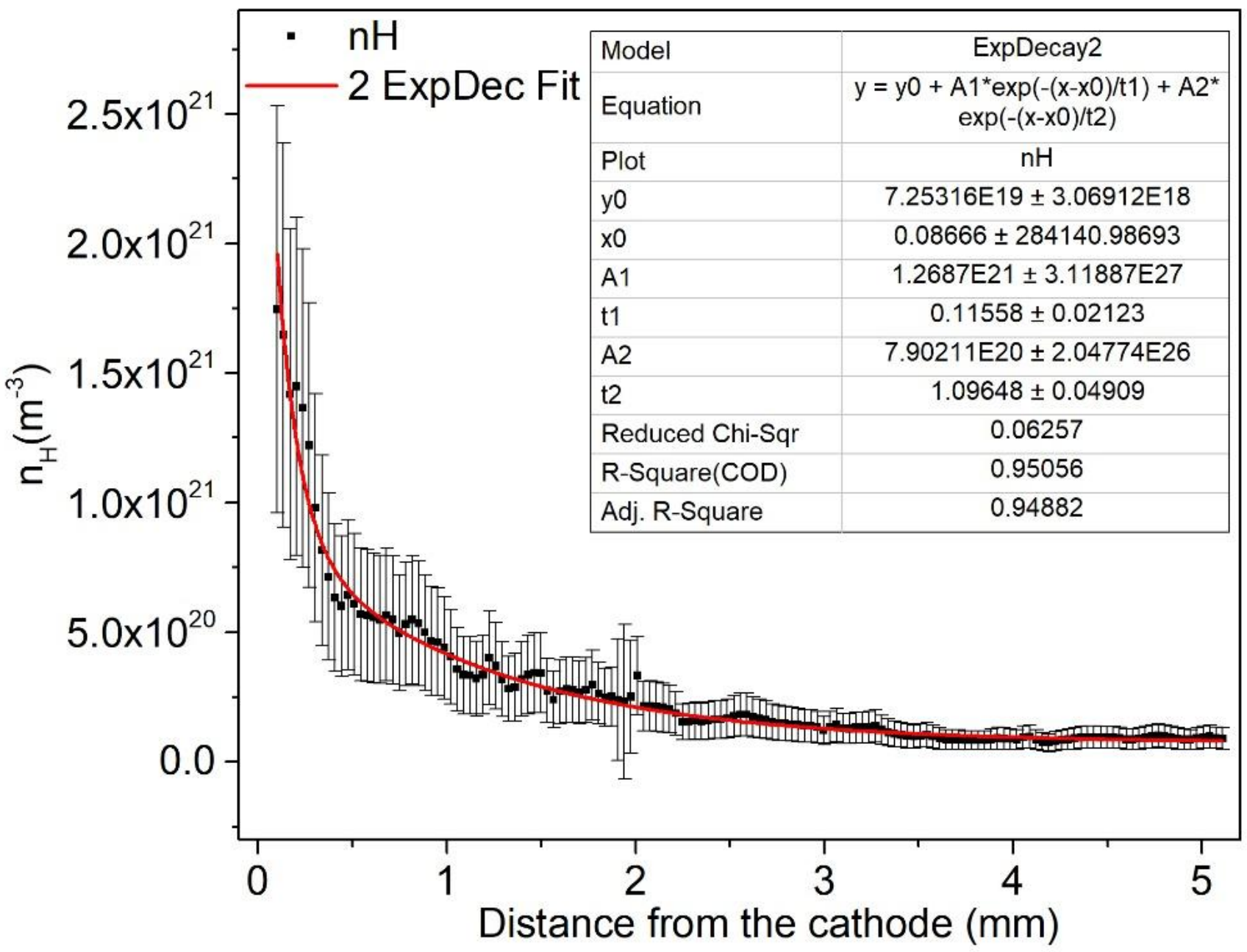

Figure 15. $n_H$ density in a cathode-adjacent zone with a 2-exponential decay fit curve.

Beyond ~0.2 mm the density decays on the longer reaction-diffusion length $t_2 \approx 1.1$ mm $\sqrt{\frac{D}{\nu}}$, where $D \approx 0.4$ $m^2/s$ is derived from low-energy H–Ar momentum-transfer cross sections ($Q_m \approx 4\text{-}6 \times 10^{-20}$ $m^2$)[82]. The effective loss rate $\nu \approx 3\text{-}4 \times 10^5$ $s^{-1}$ (lifetime ~3 μs) reflects the combined action of radical–radical and abstraction reactions in the radical-rich sheath environment. At the measured H densities (~$10^{21}$ $m^{-3}$), three-body recombination $H+H+M \rightarrow H_2 + M$ ($k_3 \approx 10^{-44}$ $m^6/s$ at 570 K) and abstraction from residual $CH_4$ or $CH_x$ fragments both contribute to H loss. While reactions such as $H+CH_2 \rightarrow CH+H_2$ ($k \approx 4.85 \times 10^{-17}$ $m^3s^{-1}$)[89] are fast, they require $CH_2$ densities of ~$10^{21}$ $m^{-3}$ to dominate the observed loss rate- too high for this transient intermediate. Thus, the observed decay length is primarily set by the spatial scale of the production zone and subsequent diffusive relaxation, rather than by a single dominant loss reaction. The double-exponential form therefore arises naturally from a narrow, cathode-adjacent production zone whose width is set by the non-linear field profile, followed by diffusion-mediated transport deeper into the discharge.

**Zone 2: the sheath-negative glow transition ($y \approx 4$-8 mm). The polymerization region.**
This extended transitional region (with peak activity centered at $y \approx 4$-6 mm) is where the chemistry shifts decisively from fragmentation to C-C bond formation and higher-hydrocarbon growth. The secondary radicals CH and, most prominently, $C_2$ exhibit broad maxima here, with the $C_2$ density (~$5.6\times10^{17}$ $m^{-3}$ at peak) exceeding that of CH (~$1.5\times10^{15}$ $m^{-3}$) by a factor of ~370 and spatially offset by several mm from the atomic H maximum (Figure 12).

The large $C_2$/CH density disparity arises from the vastly different reactivities and lifetimes of these two radicals. Methylidyne (CH) is among the most reactive neutral species in hydrocarbon plasmas, undergoing near-gas-kinetic insertion reactions with $CH_4$ ($CH+CH_4 \rightarrow C_2H_4 + H$, $k \approx 1\times10^{-16}$ $m^3$ $s^{-1}$)[89] and fast abstraction with $H_2$ ($CH+H_2 \rightarrow CH_2+H$, $k \approx 1.15\times10^{-17}$ $m^3$ $s^{-1}$)[89]. Although the local $CH_4$ and $H_2$ densities are not directly measured in this work, even taking conservative depleted estimates ($n_{CH4}$, $n_{H2}$ ~ $10^{20}$-$10^{21}$ $m^{-3}$) yields a CH loss rate of order $10^4$-$10^5$ $s^{-1}$, corresponding to lifetimes ~10-100 μs. In contrast, $C_2$ is comparatively unreactive toward closed-shell molecules[90] and is lost primarily by diffusion to the walls ($\nu_{diff}$ ~ $10^3$ $s^{-1}$, lifetime ~1 ms). The resulting ~10-100× lifetime difference accounts for the large $C_2$/CH ratio. The remaining factor of ~7 likely reflects more efficient $C_2$ production: electron-impact dissociation of accumulated $C_2H_2$ and $C_2H$ (the dominant $C_2$ precursors) may proceed faster than CH production from the increasingly depleted $CH_4/CH_3$ pool in this downstream region.

The dominant $C_2$ precursors are $C_2H_x$ species. The barrierless radical recombination reactions $CH_2 + CH_2 \rightarrow C_2H_2 + H_2$ ($k \approx 7\times10^{-20}$ $m^3s^{-1}$)[89] and $CH_3 + CH_2 \rightarrow C_2H_4 + H$ ($k \approx 7.1\times10^{-17}$ $m^3s^{-1}$)[89] efficiently convert the dense $CH_3/CH_2$ pool generated upstream into $C_2H_x$ intermediates. Subsequent electron-impact dissociative excitation of $C_2H_2$ and $C_2H$ then produces $C_2$. Thus, the sheath–negative glow boundary functions as the primary polymerization zone, where radical recombination initiates C-C bond formation and sets the stage for higher hydrocarbon growth.

**Zone 3: the negative glow (8-13.5 mm).**
In the extended negative glow, the electric field is weak, but energetic electrons are still present. Thermal atomic hydrogen persists, reducing from ~$0.8\times10^{20}$ $m^{-3}$ to ~$0.6\times10^{20}$ $m^{-3}$ between y=8 to ~13.5 mm. A distinct EISH (~3.5 eV kinetic energy) dominates the $H_\alpha$ profile. This EISH population, produced by electron impact dissociation of $CH_4$, can drive the abstraction from the $CH_4$ gas at rates much higher than thermal H (see Section 4.4). The nascent vibrational temperature exhibits a pronounced, broad re-heating peak at approximately 12.5 mm.
An additional channel for $CH_4$ dissociation in the negative glow involves Penning processes with argon metastables (Ar, $3P_0$ and $3P_2$ states at 11.5–11.7 eV). These long-lived species ($\tau$ ~ ms at these pressures) can carry excitation energy non-locally from the high-field sheath region into the field-free bulk, where electron energies are insufficient for direct dissociative excitation. The Penning dissociation reaction $Ar^* + CH_4 \rightarrow Ar + CH_3 + H$ (and related channels) proceeds with near-gas-kinetic rates ($k$~$10^{-16}$ $m^3s^{-1}$) and may contribute to the sustained H production observed in Zone 3. Quantification of this pathway requires knowledge of the Ar* density, which will be addressed in future work via laser-induced fluorescence of the Ar 4s states.
The relaxation of the chemistry in this central region is mirrored by the vibrational temperature, which decreases steadily to a minimum of ~3300 K. Here, the beam of fast electrons injected from the sheath loses energy through inelastic collisions, shifting the EEDF toward intermediate electrons (~5-15 eV) that drive less vibrational heating.

**Zone 4: anode-adjacent region ($y \gtrsim 13.5$ mm): secondary re-activation zone.**
Near the anode, a fourth chemically distinct zone emerges. After reaching a shallow minimum at y≈13-14 mm (~$1.5\times10^{20}$ $m^{-3}$), the thermal H density rises gently to ~$5.5\times10^{20}$ $m^{-3}$ at y = 16.5 mm and then sharply to ~$10^{21}$ $m^{-3}$ immediately adjacent to the anode. This anode-adjacent reactivation is attributed to the development of an anode sheath/pre-sheath, where the ambipolar field locally increases the electron temperature, re-accelerating vibrational excitation and producing a secondary source of atomic hydrogen via vibrationally-assisted dissociation. It is also possible that the ballistic ReSH "survivors" (~90 eV) identified in the transport model contribute to this zone. Although the ReSH flux in the anode region remains unquantified, such neutral bombardment could potentially supplement local field-driven dissociation by facilitating the desorption of carbonaceous adsorbates from the electrode surface. The symmetry of the discharge - intense cathode-driven chemistry mirrored by weaker but clearly detectable anode-driven re-activation - completes the millimeter-scale spatial structuring and explains the unexpected persistence of non-equilibrium conditions throughout the entire inter-electrode gap.
This reactivation is corroborated by the sharp rise in $T_{vib}$ to ~4700 K near the anode (y ≈ 16 mm). The formation of a localized positive anode sheath acts upon the bulk electron population, shifting a significant fraction above the 12.2 eV threshold required for dissociative excitation. Even though these electrons possess lower individual energies compared to the cathode beam, their higher number density results in a surge of CH(A) production with high vibrational temperatures, mirroring the excitation conditions of the cathode sheath.

### 4.4. The importance of suprathermal H atoms

The suprathermal hydrogen atoms are not just spectroscopic features- they may also drive methane activation. Morton and Kaiser[91] demonstrated computationally, that photolytically generated suprathermal H atoms exhibit significantly elevated rate constants for H + $CH_4$ abstraction in the low-temperature atmospheres of Jupiter, Saturn, and Titan, where the ambient temperature renders the thermal H population nearly chemically inert toward $CH_4$.In the short glow discharge it is the structured electron energy landscape that generates the suprathermal population, rather than photodissociation but a similar rationale extends. EISH (~17.5 eV at birth, decaying to ~3–4 eV in Zone 3) and ReSH (~110 eV, decaying to ~35–48 eV) play complementary roles: EISH provides the dominant abstraction channel because its more populous, while ReSH transports energy non-locally and vibrationally excites $CH_4$ for H abstraction.

**EISH as a chemical initiator**
The hydrogen abstraction reaction $H + CH_4 \rightarrow CH_3 + H_2$ has an activation barrier $E_a$ ≈0.516 eV[87], which is too high for thermal H at $T_g$~570 K. From the standard Arrhenius expression for the 300–2500 K range[87] , $k_{th}(570\ K) \approx 3.55\times10^{-21}\ m^3s^{-1}$. With $n_H \approx (0.6 - 0.87)\times10^{20}\ m^{-3}$ from TALIF in Zone 3 (y=7.5 - 13 mm) and $n_{CH_4} \approx 1\times10^{21}\ m^{-3}$ (feedstock value taken as an upper bound; local $CH_4$ is depleted by dissociation but not directly measured in this work), the thermal volumetric rate is:

$$R_{th}=k_{th}\cdot n_H\cdot n_{CH_4} \approx (2.1 - 3.1)\times10^{20}\ m^{-3}\ s^{-1}$$

The EISH atoms have KE ~4 eV in this zone, well above the barrier, so abstraction proceeds at the cross-section-limited rate with[88] $\sigma_{abstr} \approx 1\times10^{-20}\ m^2$. The corresponding mean free path

$$\lambda_{abst}=1/(\sigma_{abstr}\cdot n_{CH4}) \approx 10\ cm$$

is much larger than the discharge gap. EISH atoms, therefore, may bounce off the electrodes before abstracting or acquire an off-axis (radial) trajectory component. Their steady-state contribution to chemistry is set by their production rate $\Gamma_{e\text{-}dissoc} \approx 1.6\times10^{22}$ $m^{-3}s^{-1}$ (Appendix 3), which defines 2 limits. The ceiling assumes every produced EISH atom eventually abstracts before being lost:

$$R_{EISH}^{ceiling} \approx \Gamma_{e\text{-}dissoc} \approx 1.6\times10^{22}\ m^{-3}s^{-1}$$

The floor uses only the single-pass abstraction probability. For an EISH atom born mid-gap (y≈8 mm), the geometric single-pass probability is path/$\lambda_{abstr}$ ≈ 8%. Discarding any subsequent contribution from electrode reflection, multi-bounce, and off-axis (radial) trajectories:

$$R_{EISH}^{floor} \approx 0.08\cdot\Gamma_{e\text{-}dissoc} \approx 1.28\times10^{21}\ m^{-3}\ s^{-1}.$$

The resulting bracket on the suprathermal advantage is:

$$R_{EISH}/R_{th} \approx 4\text{-}6 \quad \text{(floor, single-pass only)}$$

$$R_{EISH}/R_{th} \approx 50\text{-}75 \quad \text{(ceiling, full consumption)}$$

Both ratios scale with $n_{CH_4}$ and should therefore be read as upper estimates, but the EISH advantage persists qualitatively under local $CH_4$ depletion. Even at the most conservative floor, EISH drives methane abstraction faster than the thermal H population. The realistic value lies above the floor: EISH atoms bounce off electrodes and traverse off-axis paths longer than the gap, retaining energies above the 0.5 eV barrier before thermalizing. Two caveats apply: (i) $\sigma_{abstr}$ at ~4 eV carries a factor 2–3 uncertainty; (ii) the floor neglects all post-reflection chemistry. Despite these, the result indicates that standard Arrhenius kinetics based on $T_g$ substantially underestimate methane reactivity in structured low-pressure discharges. The suprathermal EISH population provides a non-equilibrium pathway to bypass the activation barrier in regions where the non-local EEDF sustains its production.

**ReSH as energy carrier and vibrational primer**

The ReSH population (~110 eV) acts as a secondary agent of non-local energy transport. Based on a total ion flux to the cathode of $\gamma_{ion}$~$4.6\times10^{15}$ $s^{-1}$ and branching parameters detailed in Appendix 3, the total power transported by this ballistic cohort is estimated to be:

$$P_{ReSH} \sim \gamma_{ion}\times(0.5\text{-}0.7)\times(0.3\text{-}0.5)\times(0.05\text{-}0.2)\times 110\ eV \sim 0.6\text{-}5.7\ mW$$

This represents approximately 0.3-3 % of the total electrical power dissipated by the plasma (0.19 W = $I_d\times V_p$=0.736 mA×264 V). While their contribution to bulk gas heating is minor compared to the EISH population, these hyperthermal atoms still influence local methane dynamics in Zone 2 through non-reactive channels.

Beyond energy transport, ReSH atoms interact with $CH_4$ in Zone 2 through several collision channels. Trajectory scaling from hyperthermal Ar + $CH_4$ analogues shows that direct bond-breaking channels such as Collision-Induced Dissociation ($\sigma_{CID}$ ~$0.1\text{-}0.5\times10^{-20}$ $m^2$)[92] or hyperthermal abstraction ($\sigma_{abs}$≈$0.05\text{-}0.08\times10^{-20}$ $m^2$)[88] - are suppressed by the unfavorable H/$CH_4$ mass ratio and remain negligible. Instead, translation-to-vibration (T-V) energy transfer forms the dominant inelastic channel ($\sigma_{vib}$~$5\times10^{-20}$ $m^2$), depositing 1-5 eV into methane internal modes per collision. The calculated ReSH flux, $\Phi_{ReSH}$~$3\times10^{16}$-$3\times10^{17}$ $m^{-2}s^{-1}$, follows from the production-rate integration of Appendix 3 or equivalently from the ion current density $J_i \approx 0.77$ A/m² combined with the $CH_x^+$ fraction (0.5-0.7), reflection coefficient ($R_n$ = 0.3-0.5), and H branching (0.05-0.2). Using the upper-end value with $n_{CH_4} \sim 10^{21}$ $m^{-3}$ (feedstock upper bound, as in the EISH analysis):

$$R_{ReSH} \approx \Phi_{ReSH}\times n_{CH4}\times\sigma_{vib} \sim 10^{18}\text{-}10^{19}\ m^{-3}s^{-1}$$

The absolute rate is modest, but the spatially localized "vibrational priming" softens the C-H bonds within the transition zone, effectively reducing the barrier for subsequent thermal-H abstraction. Experiments on H + $CH_4(v_3)$ and similar H + $CHD_3(v_1)$ reactions show that exciting the C-H stretch by one quantum raises the abstraction rate by an order of magnitude or more[93–95].

**5. Conclusions**

This study provides a comprehensive, spatially resolved investigation of methane decomposition within a short, low-pressure DC glow discharge in an Ar-$CH_4$ mixture. The distributions of key reactive species (H, CH, $C_2$) were mapped and correlated with the discharge's unique, non-uniform structure. The main conclusions are as follows:

1. Dual dissociation mechanisms: High-resolution spectroscopy of the Hα emission line revealed the presence of two distinct populations of suprathermal hydrogen atoms. A very energetic population (~110 eV) is attributed to the reflection of hydrocarbon ions from the cathode surface, while a second, more populous group (~17.5 eV) is produced by direct electron-impact dissociation of methane in the sheath volume. This provides direct evidence for the parallel operation of both surface- and volume-based dissociation mechanisms. The spatial evolution of the ReSH component provides experimental evidence of kinetic “velocity filtration” in a collisional discharge environment. The anode region reveals a distinct ballistic cohort (~90 eV) that traversed the gap while maintaining axial directionality invisible to the side-on measurement, demonstrating that the observed KE profile represents a visibility-weighted superposition of scattered and ballistic populations.

2. Suprathermal chemical initiation: The EISH population may act as a significant methane activator beyond the sheath. Kinetic estimates reveal that these high-energy atoms, overcoming activation barriers, drive the production of methyl radicals ($CH_3$) at a rate of $\geq$ 4-6 times faster than the entire bulk thermal hydrogen population. This result challenges the reliance on thermal rate coefficients for modeling low-pressure hydrocarbon plasmas.

3. Spatially segregated chemistry: The short glow discharge is segregated into distinct chemical zones. Methane decomposition is confined to the cathode sheath, while polymerization is delayed to the sheath-negative glow boundary. This downstream zone is dominated by $C_2$ radicals ($n_{C2} \gg n_{CH}$), a disparity attributed to the selective quenching of polar CH radicals in the ion-rich transition region. The discharge symmetry is completed by a weaker but detectable re-activation zone near the anode, likely driven by local electron acceleration and possibly supplemented by ballistic neutral bombardment.

4. Polymerization frontier: The densities of the secondary radicals CH and $C_2$ peak sharply at the sheath-negative glow boundary. The peak density of $C_2$ (~$5.6\times10^{17}$ $m^{-3}$) was found to be two orders of magnitude greater than that of CH (~$1.5\times10^{15}$ $m^{-3}$), indicating efficient pathways leading to C-C bond formation in this region. This zone acts as a "polymerization region" where the primary fragments generated in the sheath accumulate and react.

5. Non-equilibrium chemistry: The plasma operates far from thermodynamic equilibrium. While the bulk gas temperature remains cool and relatively uniform at approximately 570 K, the emitting CH(A) radicals exhibit an extremely high vibrational temperature (3300–4700 K). This disparity is not a measure of the bulk gas energy but serves as a direct fingerprint of the high-energy dissociative excitation channel ($e^- + CH_4 \rightarrow CH(A) + H_2 + H$) that creates them, confirming that the chemistry is driven by specific, non-thermal electron-impact processes.
6. The axially resolved density profiles of H, CH, and $C_2$ radicals offer essential benchmark data for validating 1D and 2D plasma kinetic models, especially those accounting for non-local electron energy distribution functions in structured discharges like this short GD setup.

These findings demonstrate how the non-local electron kinetics inherent to a short glow discharge can be leveraged to spatially control chemical reactions. The detailed species maps provide critical benchmark data for the validation of kinetic models and offer a foundational understanding for optimizing plasma-based methane conversion technologies.

Future work will extend these insights by varying the $CH_4$ fraction (e.g., 5-30% in Ar) to assess its influence on dissociation efficiency and radical yields and incorporating plasma potential measurements.

Finally, future efforts must prioritize the direct, quantitative measurement of suprathermal hydrogen populations to fully elucidate their role in the discharge. Specifically, determining the absolute density of these "hot" atoms is crucial for quantifying their impact on methane decomposition chemistry, while resolving their angular distribution is essential for understanding the underlying physics of non-isotropic molecular dissociation[9,83]. In this context, the ground-state H-TALIF absorption scan at y = 1.8 mm (Figure *3*.b) potentially provides intriguing preliminary evidence; the subtle intensity enhancements observed at approximately ~±40 pm from the line center align with the Doppler shift expected for the ~17.5  eV EISH population identified in emission. However, given that these features are near the noise floor and constrained by the OPO system's ~10 pm tuning limit and ~12.3 pm laser bandwidth, they remain a tentative suggestion of ground-state EISH detection. This observation defines a clear path for future high-resolution studies; the use of narrow bandwidth lasers to resolve these ground-state velocity wings would allow for the measurement of the suprathermal flux and its directional anisotropy, providing a foundational benchmark for the next generation of non-local kinetic plasma models.

**Acknowledgement**

This work is supported by the Princeton Collaborative Research Facility (PCRF), which is supported by the U.S. Department of Energy (DOE) under Contract No. DE-AC02-09CH11466.
The author is grateful to Dr. Yevgney Raitses, Dr. Nirbhav Chopra, Dr. Mikhail Mokrov for helpful discussions and to Mr. Tim Bennet for technical assistance.

**Appendix 1: vibrational and rotational temperatures in calculations of CH and $C_2$ Boltzmann fractions**

In non-equilibrium plasmas like the short glow discharge studied here, the vibrational ($T_{vib}$) and rotational ($T_{rot}$) temperatures of radicals such as CH and $C_2$ often deviate from the gas temperature ($T_g \approx 570$ K) due to production mechanisms that impart excess energy and relaxation timescales that exceed the species' residence time. This appendix discusses the use of $T_{vib} \approx$ 3300-4700 K (from CH(A-X) emission fits) and $T_{rot}$ = 570 K for the Boltzmann fraction ($f_B$) in LIF density calculations for both CH(X) and $C_2$(a).

In non-equilibrium plasmas, the vibrational ($T_{vib}$) and rotational ($T_{rot}$) temperatures of radicals often deviate from the bulk gas temperature ($T_g \approx$ 570 K)[96]. Accurate density measurements via LIF require a realistic estimate of the Boltzmann fraction ($f_B$) of the probed state. This appendix details the justification for using $T_{rot} \approx T_g$ (equilibrated) while assuming a non-equilibrium $T_{vib}$ based on the competition between relaxation and diffusion timescales.

### A.1 Characteristic radical lifetimes (diffusion vs flow)

The time available for a radical to equilibrate with the bulk gas is limited by its loss mechanisms. The residence time in plasma is defined by the volumetric gas flow in the chamber and the volume of plasma (occupied by CH/$C_2$). The plasma volume can be calculated from plasma images (similar to the plasma diameter estimation at Figure 5b) to be $V_{plasma} \approx 15.4$ cm$^3$. The total gas flow into the chamber is $Q_{std}$ =35 sccm (35 cm$^3$/minute). Adjusting the $Q_{std}$ to the true volumetric flow rate:

$$\mathrm{Q_{chamber}} = \mathrm{Q_{std}} \cdot \frac{\mathrm{P_{std}}}{\mathrm{P_{chamber}}} \cdot \frac{\mathrm{T_g}}{\mathrm{T_{std}}} = 35\ cm^3/min\ \times \frac{760\ Torr}{0.4\ Torr} \times \frac{570\ K}{273\ K} \approx 138445 \frac{cm^3}{min}$$

The residence time then is $\tau_{res} = \frac{V_{plasma}}{Q_{chamber}} \approx 6.65\ ms$.

However, for reactive radicals like CH and $C_2$, diffusive loss to the walls is significantly faster. Using the hard-sphere approximation at 400 mTorr and 570 K in the Ar-dominated mixture, the diffusion coefficients are estimated as $D_{CH} \approx 940\ cm^2/s$ and $D_{C_2} \approx 700\ cm^2/s$.

The characteristic diffusion time $\tau_{diff} = \Lambda^2/D$, where $\Lambda$ is the characteristic diffusion length scale. Considering a range of scales from sharp local gradients ($\Lambda$~0.1 mm) to the sheath boundary ($\Lambda$~5 mm), the diffusion lifetimes range from 0.1 to 350 µs. For the bulk measurements reported here, the upper end of this range (~200-350 µs) is shorter than the gas residence time.

### A.2 Equilibration in the excited CH(A) state

The CH($A^2\Delta$) state, observed in emission, is primarily produced via electron-impact dissociative excitation of $CH_4$ (e + $CH_4 \rightarrow$ CH(A) + H + $H_2$ + e, threshold ~15.5 eV), which channels excess energy (~5–10 eV) into high vibrational and rotational excitation in CH(A), yielding nascent $T_{vib}$ and $T_{rot}$ ~3300–4700 K as observed in emission.

To assess equilibration before radiative decay, the collision frequency $\nu = n_g \sigma v_{rel}$ is calculated.

- $n_g \approx 6.78 \times 10^{21}\ m^{-3}$ (at P=400 mTorr, $T_g$=570 K)
- $\sigma \approx 5\times10^{-19}$ m² (conservative for excited diatom-Ar collisions)
- $v_{rel} = \sqrt{\frac{8k_BT_g}{\pi\cdot\mu}}$ =1105.1 m/s, where $\mu \approx 1.64\times10^{-26}$ kg is the reduced mass of CH-Ar.

Thus, $v_{rel} \approx 3.75\times 10^6$ s$^{-1}$, and the time between collisions $\tau_{coll} \approx 0.267$ µs.

The radiative lifetime of CH(A) is $\tau_{rad} \approx 0.53$ µs, CH(A) decays to CH(X) after ~2 collisions on average.

### A.3 Equilibration in ground CH(X) state

CH(A) decays radiatively to CH(X), preserving $T_{vib} \approx 3300–4700$ K in CH(X). The linear Boltzmann plot from CH-LIF (Section 3.3, pumping R2(1), R2(4), R2(7)) indicates a single $T_{rot} \approx 570$ K, consistent with fast R-T relaxation. The equilibration depends on Relaxation Energy Transfer (RET) and Vibrational Energy Transfer (VET) rates compared to the diffusion window (>200-350 µs for bulk).

- RET: In CH(X), $k_{RET} \approx 10^{-16}$ cm³ s$^{-1}$, so $\tau_{RET}=1/(k_{RET} \cdot n_g) \approx 1.5$ µs. This ensures $T_{rot}$ equilibrates to $T_g \approx 570$ K, within the residence time, as observed.
- VET: For Ar (85%): $k_{VET}(Ar) \approx 0.5–6.8\times10^{-20}$ m$^3$s$^{-1}$, $n_{Ar} \approx 5.76 \times 10^{21}$ m$^{-3}$. For $CH_4$ (15%): $k_{VET}(CH_4) \approx 10^{-18}$ m³s$^{-1}$ (estimated from analogous radicals like OH), $n_{CH_4} \approx 1.02 \times 10^{21}$ m$^{-3}$. Total $\tau_{VET} \approx 0.71–0.95$ ms > $\tau_{diff}$).

Since $\tau_{VET} < \tau_{diff}$ (for bulk scales), most “hot” CH radicals are removed from the plasma before they can vibrationally equilibrate. They retain a "memory" of their nascent $T_{vib}$.

### A.4 Application to $C_2(a^3\Pi_u)$ state

The $C_2$(a) is formed via secondary reactions, e.g., CH + CH → $C_2$+$H_2$ (~4–5 eV exothermic) or C+CH→$C_2$ + H, yielding nascent $T_{vib}$ ~3000–4000 K. Spatial density peaks near the sheath-NG transition (y≈4–5 mm) suggest similar chemistry to CH.

- RET: Equivalent to CH, $\tau_{RET}=1/(k_{RET} \cdot n_g) \approx 1.5$ µs ($\ll \tau_{res}$), sufficient for thermalization.
- VET: $k_{VET}(Ar) \approx 10^{-19}–10^{-18}$ m$^3$s$^{-1}$ (estimated from diatomic-Ar studies), $k_{VET}(CH_4) \approx 10^{-18}$ m$^3$s$^{-1}$. Total $\tau_{VET} \approx 150–628$ µs.

Therefore, $C_2$(a) also retains a non-equilibrium $T_{vib}$.

### A.5 Direct production of CH(X) with “cold” $T_{vib}$

The possibility of production of CH(X) by alternative pathways with lower $T_{vib}$, needs to be evaluated. The “hot” CH(A) comes from e + $CH_4$ → CH(A) + H + $H_2$ + e, subsequently decaying to CH(X) that largely maintains “hot” $T_{vib}$. Energetic threshold for this reaction is ~12 eV (peak cross section for 50-100 eV), requiring energetic electrons abundant in the cathode sheath and sheath-NG boundary (sheath EEDF has ~10-30% beam electrons >10-15 eV[97,98]) and cross-section is $\sigma \sim 2\times10^{-21}$ m$^2$ (from dissociative excitation models)[70].

The "cold" pathways produce ground-state CH(X) directly or stepwise, with lower excess energy leading to lower nascent $T_{vib}$ (~1000–1500 K, as estimated from low-excess-energy partitioning). Key reactions are:

Direct: e+$CH_4$→CH(X)+$H_3$+e (threshold ~12 eV).

Stepwise: 1) e+$CH_4$→$CH_3$+H+e (threshold ~8.8 eV), then

2) e+$CH_3$→CH(X)+$H_2$+e (threshold ~4.3 eV).

3) Secondary reactions like $CH_2$+ $CH_2 \rightarrow CH(X)+CH_3$

Literature indicates that in methane-containing discharges the non-equilibrium "hot" population accounts for approximately 20% to 35% of the nascent distribution depending on the specific gas mixture, with the remaining fraction relaxing to the thermalized "cold"' background[63,64]. This empirical distribution profile is adapted in the present work to compute the Boltzmann partition fractions ($f_B$) for the CH and $C_2$ ground states.

**Appendix 2: Cross-section estimations**

This appendix provides detailed derivations for the cross sections marked with "*" in Table 2, which were estimated indirectly due to limited direct experimental data. Estimations are based on scaling from analogous reactions, theoretical models (e.g. Polarized Trajectory (PT) methods incorporating ion-induced dipole polarization[99,100], Collision-Induced Dissociation (CID) models for fragmentation channels[9,83], Gloumousis-Stevenson[101] (ion-molecule collisions) orbiting theory or polarization potentials[102] (ion-neutral) ), and empirical adjustments for energy dependencies, exothermicity/endothermicity, and molecular properties like polarizability ($\alpha$) and reduced mass ($\mu$). Units are in m², and values are applicable for ion energies around 50–200 eV unless noted. Citations refer to experimental or theoretical sources where available; uncertainties are typically 20–50% for scaled values.

- **$CH_3^+$+$CH_4$ CEX** ($\sigma \approx 2\times10^{-20}$ m² at ~ 200 eV): This charge exchange (CE) reaction, $CH_3^+ + CH_4 \rightarrow CH_3 + CH_4^+$, is endothermic by $\Delta \approx$ IP($CH_4$) - IP($CH_3$) = 12.6 eV-9.84 eV=2.76 eV. The base value is scaled from the measured proton transfer $CH_4^+ + CH_4 \rightarrow CH_5^+ + CH_3$ ($\sigma \approx 1\text{-}2\times10^{-19}$ m² at 10 eV, decreasing to ~$5\times10^{-20}$ m² at 200 eV via $1/\sqrt{E}$)[49]. Scaling up by ×0.4 for non-resonant CE and higher E; following the Gloumousis-Stevenson orbiting model, where $\sigma \propto \sqrt{(\alpha/\mu)}$ with $\alpha_{CH_4} \approx 2.6$ Å³ - yields ~$2\times10^{20}$ m².

- **$CH_5^+$+$CH_4$ CEX** ($\sigma \leq 5\times10^{-21}$ m² at ~200 eV): $CH_5^+ + CH_4 \rightarrow CH_5 + CH_4^+$ (strongly endothermic $\Delta \approx$ +5.6 eV; IP($CH_4$) ≈ 12.6 eV, IP($CH_5$) ≈ 7 eV). Colgate & Schmidt (1966)[49] measured proton transfer $CH_4^+ + CH_4 \rightarrow CH_5^+ + CH_3$, not reverse CEX. Langevin[99] $\sigma_L \approx 1.6\times10^{-19}$ m² scaled by ×0.01 for high-$\Delta$ non-resonant suppression (empirical from $\Delta \approx 4$ eV systems; Anicich[100]) yields $\sigma \leq 5\times10^{-21}$ m² at 200 eV.

- **$CH_3^+$+Ar CEX** ($\sigma \approx 4\times10^{-21}$ m² at ~200 eV): Reaction $CH_3^+ + Ar \rightarrow CH_3 + Ar^+$ is exothermic by $\Delta \approx$ IP(Ar) - IP($CH_3$) = 15.76 eV - 9.84 eV = 5.92 eV. Scaled from $CH_4^+$+Ar proton abstraction ($\sigma \approx 5\times10^{-21}$ m²)[50], via similar mass and $\Delta$. High-$\Delta$ non-resonant suppression (Rapp & Francis, 1962)[51] supports ×0.1 from resonant analogs.

- **$CH_5^+$+Ar CEX** ($\sigma \approx 4\times10^{-21}$ m² at ~200 eV): $CH_5^+ + Ar \rightarrow CH_5 + Ar^+$ (exothermic $\Delta \approx 8.76$ eV; IP(Ar) = 15.76 eV, IP($CH_5$) ≈ 7 eV). No direct data. Scaled from measured $CH_4^+$ + Ar CE ($\sigma \approx 5\times10^{-21}$ m² at 200 eV; Peko et al., 1998)[50] via higher $\Delta$ and mass. High-$\Delta$ non-resonant suppression (Rapp & Francis, 1962)[51] supports ×0.8 from $CH_4^+$ case.

- **$Ar^+$+$CH_4$ elastic** ($\sigma \approx 1.5\times10^{-19}$ m² at ~1 eV): At low energy, Langevin ion-induced dipole capture $\sigma_L = 2\pi e\sqrt{(\alpha/\mu)} \approx 1.5\times10^{-19}$ m² ($\alpha_{CH_4} \approx 2.6$ Å³, $\mu \approx 20$ amu). Energy-independent below ~10 eV. No direct data (Peko[50] starts at ~5 eV).

- **$Ar^+$+$CH_4$ elastic** ($\sigma \approx 3\times10^{-20}$ m² at ~200 eV): Interpolated from $Ar^+$ + $H_2$ elastic ($\sigma\approx1.2\times10^{-19}$ m² at 100 eV)[66], scaled ×2.5 for $CH_4$ polarizability ($\alpha_{CH_4} \approx 2.6$Å³ vs. $\alpha_{H_2} \approx 0.8$ Å³) using hard-sphere model $\sigma \propto (r_{ion} + r_{target})^2$. Polarization potential $V=-\alpha\cdot e^2/(2\cdot r^4)$ gives orbiting $\sigma \approx \pi (\alpha\cdot e^2\cdot/E)^{1/2}$~$2.5\times10^{-20}$ m² at 200 eV. Peko's[50] total σ (CE + CID) ≈ $2\times10^{-20}$ m² supports elastic dominance.

- **$CH_3^+$+$CH_4$ elastic** ($\sigma \approx 1.8\times10^{-19}$ m² at ~1 eV): At low energy, Langevin ion-induced dipole capture $\sigma_L = 2\pi e\sqrt{(\alpha/\mu)} \approx 1.8\times10^{-19}$ m² ($\alpha_{CH_4} \approx 2.6$ Å³, μ ≈ 7.5 amu). Energy-independent below ~10 eV. No direct data (Peko[50] starts at ~5 eV).

- **$CH_3^+$+$CH_4$ elastic** ($\sigma \approx 2\times10^{-20}$ m² at ~200 eV): Ion-induced dipole $\sigma \approx 2\pi\sqrt{(\alpha e^2/\mu E)}$~$1.8\times10^{-20}$ m² (μ≈ 7.5 amu, α ≈ 2.6 Å³). Scaled from $CH_4^+$+$CH_4$ symmetric ~$2.5\times10^{-20}$ m² (Anicich, 1993)[103]. Peko's[50] total σ ≈ $5\times10^{-20}$ m² supports elastic ~40%.

- **$CH_4^+$+$CH_4$ elastic** ($\sigma \approx 2.0\times10^{-19}$ m² at ~1 eV): At low energy, Langevin ion-induced dipole capture $\sigma_L = 2\pi e\sqrt{(\alpha/\mu)} \approx 2.0\times10^{-19}$ m² ($\alpha_{CH_4} \approx 2.6$ Å³, μ ≈ 8 amu). Energy-independent below ~10 eV. No direct data (Peko[50] starts at ~5 eV).

- **$CH_4^+$+$CH_4$ elastic** ($\sigma \approx 3\times10^{-20}$ m² at ~200 eV): Symmetric; polarization model $\sigma\approx\pi\sqrt{(2\alpha e^2/E)}$ ~$3.5\times10^{-20}$ m² ($\alpha_{CH_4} \approx 2.6$ Å³). Peko's[50] total σ ≈ $5\times10^{-20}$ m² supports elastic ~60%.

- **$CH_5^+$+$CH_4$ elastic** ($\sigma \approx 2.0\times10^{-19}$ m² at ~1 eV): At low energy, Langevin ion-induced dipole capture $\sigma_L = 2\pi e\sqrt{(\alpha/\mu)} \approx 2.0\times10^{-19}$ m² ($\alpha_{CH_4}\approx2.6$ Å³, μ ≈ 9.5 amu). Energy-independent below~10 eV. No direct data. Analogous $H_3^+$ + $H_2$ ~$1.8\times10^{-19}$ m² (Anicich[100])

- **$CH_5^+$+$CH_4$ elastic** ($\sigma \approx 3\times10^{-20}$ m² at ~200 eV): Scaled from $H_3^+$+$H_2$ elastic ~$2.5\times10^{-20}$ m² at 100 eV (Anicich[103]), upscaled for higher $\alpha_{CH_4}$. Polarization orbiting $\sigma \approx \pi \sqrt{(\alpha e^2/E)}$ ~$3\times10^{-20}$ m². Peko's[50] total σ for $CH_4^+$ + $CH_4$ ≈ $5\times10^{-20}$ m² supports elastic ~60%.

- **$CH_3^+$+$CH_4$ inelastic** ($\sigma \approx 5\times10^{-21}$ m² at ~200 eV): CID channels (threshold ~10 eV). Peko's[50] total σ ≈ $5\times10^{-20}$ m²; CID ~10–20% of total. Scaled from $H^+$ + $H_2$ CID ~$10^{-20}$ m² at 100 eV[104].

- **$CH_4^+$+$CH_4$ inelastic** ($\sigma \approx 1\times10^{-20}$ m² at ~200 eV): Symmetric CID to $CH_3^+$ + $CH_5$/H (threshold ~1 eV). Peko's[50] total σ ≈ $5\times10^{-20}$ m²; CID ~20% of total.

- **$CH_5^+$+$CH_4$ inelastic** ($\sigma \approx 1\times10^{-20}$ m² at ~200 eV): Proton transfer/dissociation; reverse of $CH_4^+$ + $CH_4$ → $CH_5^+$ + $CH_3$. Peko's[50] total σ ≈ $5\times10^{-20}$ m²; PT ~60% of total.

- **$CH_3^+$+Ar inelastic** ($\sigma \approx 1\times10^{-21}$ m² at ~200 eV): Minor CID/electronic; scaled from neutral analogs ×10 for charge. Peko's[50] $CH_4^+$ + Ar total σ ≈ $5\times10^{-21}$ m²; CID ~20%.

- **$CH_5^+$+Ar inelastic** ($\sigma \approx 2\times10^{-21}$ m² at ~200 eV): Threshold ~8–10 eV; scaled from $CH_4^+$+Ar CID (Peko's[50] σ ≈ $5\times10^{-21}$ m², CID ~40%).

- **H + Ar/$CH_4$ elastic** ($\sigma \approx 1.0\times10^{-19}$ m² at 110 eV; ~$1.2\times10^{-19}$ m² at 17.5 eV) Ruzic and Cohen[71] measured large-angle scattering (>20 mrad) σ ≈ $6\times10^{-20}$ m² at 110 eV. Forward glory contributes ~2/3 of total $\sigma_{el}$ → apply ×3 correction → $\sigma_{el}(Ar) = 1.8\times10^{-19}$ m². For $CH_4$ scaled by polarization ($\alpha_{CH_4} = 2.59$ Å³, $\alpha_{Ar} = 1.64$ Å³) + geometric size factor → $\sigma_{el}(CH_4) = 2.84\times10^{-19}$ m². In 85% Ar/15% $CH_4$ mix: $\sigma_{eff} = 1.94\times10^{-19}$ m².

For the 17.5 eV case, the cross section is typically higher, and $\sigma_{el}$ increases by ~20% due to the velocity-dependent Ramsauer minimum in H-Ar.

**Appendix 3. H production rates estimation: CEX vs electron-impact**

**Reflection rate:** The ion current at the cathode surface is 0.77 A/m$^2$ (Figure 5.b), corresponding to ion flux $\Gamma_i \approx 4.81\times10^{18}$ ions/m$^2$ s$^{-1}$. Assuming $CH_x^+$ comprises 50-70% of the total ions and using the active cathode area ~$9.62\times10^{-4}$ m$^2$ - the rate of $CH_x^+$ hitting the cathode is $\approx 2.32\text{-}3.24\times10^{15}$ s$^{-1}$. Factoring in reflection coefficient ($R_n$~0.3-0.5 for $CH_x^+$ on Cu)[78] and branching to H (~0.05-0.2, for $CH_5^+/CH_4^+$)[61] yields an H production rate of $\approx 3.48\times10^{13}\text{-}3.24\times10^{14}$ H/s. Volume production factors in sheath volume = $3.85 \times 10^{-6}$ m$^3$, resulting in $\Gamma_{ref} \approx 0.9\text{-} 8.4\times10^{19}$ H·m$^{-3}$s$^{-1}$.

**CEX rate:** $\Gamma_{CE} = n_i \cdot n_n \cdot \langle \sigma_{CE} \cdot v_i \rangle \cdot \eta_{dissoc} \approx 0.56\text{-}2\times10^{21}$ H·m$^{-3}$s$^{-1}$ , $n_i = 10^{14}$ m$^{-3}$ ($CH_x^+$), $n_n = n_g = 6.77\times10^{21}$ m$^{-3}$, $\sigma_{CEeff} = 5.35\times10^{-20}$ m$^2$ (weighted for Ar/$CH_4$ mix), $v_i = 1.55\times10^5$ m/s (for energetic ions in the sheath), $\eta_{dissoc} = 0.1\text{-}0.35$[50].

**Electron-impact production rate:** $\Gamma_{e\text{-}dissoc} = n_e \cdot n_{CH4} \cdot \sigma_{e_diss} \cdot v_e = 1.6\times10^{22}$ H·m$^{-3}$s$^{-1}$ , $n_e = 0.1 \cdot 3.3\times10^{15}$ m$^{-3}$ (factor 0.1 for energetic tail), $n_{CH4} = 1\times10^{21}$ m$^{-3}$, $\sigma_{e_diss} = 10^{-20}$ m$^2$, $v_e \approx 5\times10^6$ m·s$^{-1}$ (for 50-100 eV $e^-$)

**Appendix 4. Iterative 3-component $H_\alpha$ fitting:** The fitting process used a sequential approach to deconvolve the $H_\alpha$ line into three Gaussian components, with each step refining initial guesses and constraints. Matlab's lsqcurvefit function was used for curve fitting.

**Step 1: single Gaussian fit:** A single Gaussian was fitted to the normalized spectrum to model the MG component.

**Step 2: broad Gaussian fit**: The residual between the single Gaussian fit and the spectrum was calculated to capture wing contributions. A broad Gaussian (BG) was fitted to this residual, with initial amplitude estimated from the residual's maximum.

**Step 3: two-Gaussian fit**: Using parameters from the MG and BG, a two-Gaussian model (BG + MG) was fitted, with amplitude and FWHM bounds set to ±30% of the broad Gaussian's values from Step 2.

**Step 4: narrow Gaussian fit**: The residual from the two-Gaussian fit was analyzed within a ±0.02 nm window around the peak wavelength to isolate the narrow component. A narrow Gaussian was fitted with initial $FWHM_3$ = 0.018 nm and bounds ensuring $FWHM_3 \leq 0.035$ nm.

**Step 5: 3-component fitting:** All three components (BG,MG, NV) were fitted simultaneously to the spectrum, using parameters from previous steps and bounds of ±30% for amplitudes and FWHMs of the broad and narrow components. The final parameters included amplitudes ($A_1$, $A_2$, $A_3$), FWHMs ($FWHM_1$, $FWHM_2$, $FWHM_3$), and peak wavelengths ($\mu_1$, $\mu_2$, $\mu_3$). The fitting was repeated for 5 spectra recorded in succession. Mean and standard deviation of the parameters (FWHM and amplitude for each component) were calculated across spectra for each spatial position.

**Appendix 5. Kinetic modeling of ReSH transport: a 3-zone model**

**Model description and mathematical formulation:** To deconvolve the competing effects of collisional cooling and particle removal on the suprathermal hydrogen population, a one-dimensional (1D) kinetic transport model was developed. The model evolves the energy distribution function of the H atoms, N(E,y) as they traverse the inter-electrode gap from the cathode (y=0) to the edge of Zone 3 (y=13.5 mm). Evolution is governed by three coupled mechanisms that act on the distribution at each spatial step *dy*:

1. **Cooling (energy loss):** Collisions transfer kinetic energy from the H atoms to the background gas. This is modeled as a continuous shift of the energy grid. For an atom with energy $E$, the energy remaining after a step $dy$ is:

$$E(y + dy) = E(y) - n_{gas} \cdot \sigma_{cool}(E) \cdot \xi \cdot E(y) \cdot dy$$

where $n_{gas}$ is the gas density, $\sigma_{cool}$ is the effective cross-section for momentum transfer, and $\xi$ is the fractional energy loss per collision. The cooling cross-section follows a weak energy dependence:

$$\sigma_{cool}(E) = \sigma_{cool,0} \left(\frac{E_{ref}}{E}\right)^{0.25}$$

2. **Filtration (geometric loss):** Atoms are lost due to energy-dependent scattering out of the spectroscopic line-of-sight. The filtration cross-section scales inversely with velocity ($\sigma \propto \frac{1}{v}$), reflecting the higher probability of large-angle scattering for slower atoms:

$$\sigma_{filt}(E) = \sigma_{filt,0} \sqrt{\frac{E_{ref}}{E}}$$

The population density decays according to the total removal rate:

$$N(E, y + dy) = N(E, y) \cdot e^{-n_{gas} \cdot \sigma_{filt}(E) \cdot dy}$$

**Zone-specific interaction regimes:** To reproduce the complex non-monotonic evolution of the average kinetic energy, the model employs zone-dependent parameters that reflect the changing gas composition and local environment:

- **Zone 1 (cathode sheath):** $CH_4$-depleted, Ar-dominated. Low cooling rate (H-Ar elastic, small $\xi$) and low filtration losses.
- **Zone 2 (transition):** $CH_4$-rich region. Enhanced cooling cross-section and fractional energy loss due to efficient inelastic H-$CH_4$ collisions (vibrational excitation, CID). Moderate filtration.
- **Zone 3 (negative glow):** Returns to Ar-dominated composition. Cooling parameters revert to Zone 1 values, but filtration cross-section increases substantially, causing velocity-selective removal that produces the observed "beam hardening".

**Zone-specific physics and fitted parameters:** The model parameters ($\sigma_{cool}$,$\sigma_{filt}$ $\xi$) were fitted to the experimental $\langle E \rangle$ profile using a minimization algorithm. The results reveal three distinct transport zones (*Table 3*).

**Zone 1: cathode sheath (0-4 mm)**. Transport is dominated by elastic H-Ar scattering. The mass mismatch (1/40) limits energy transfer to $\xi$ ≈3%. The low filtration cross-section reflects minimal geometric losses in this region.

**Zone 2: transition (4-8 mm).** The "Kinetic Crash" is driven by inelastic H-$CH_4$ collisions. Fast H atoms transfer energy efficiently to $CH_4$ vibrational/rotational modes, with $\xi \approx 13\%$ per collision- substantially higher than in Ar. The enhanced cooling cross-section ($\sigma_{cool}$ ~6.5×10$^{-19}$ m$^2$) reflects the larger $CH_4$ molecular size and additional inelastic channels. Filtration remains moderate in this zone.

**Zone 3: negative glow (8-13.5 mm)**. The apparent re-heating is caused by velocity filtration. As atoms slow down, scattering becomes increasingly isotropic. Slower atoms are selectively removed by energy-dependent geometric scattering ($\sigma_{filt} \propto \frac{1}{\sqrt{E}}$), leaving fast survivors. The cooling parameters ($\sigma_{cool}$, $\xi$) return to Zone 1 values, confirming the Ar-dominated composition. The key difference is the substantially elevated filtration cross-section ($\sigma_{filt}$ ~12×10$^{-19}$ m$^2$), which drives the preferential removal of low-energy atoms and produces the observed rise in average energy.

Table 3.Summary of zone-dependent transport parameters. σ values refer to the effective cross-section at reference energy $E_{ref}$=10 eV.

| Zone | Region | y(mm) | Dominant process | $\sigma_{cool}$(10$^{-19}$ m$^2$) | $\sigma_{filt}$(10$^{-19}$ m$^2$) | ξ(%) | $\langle E \rangle$ (eV) |
|---|---|---|---|---|---|---|---|
| 1 | Cathode Sheath | 0-4 | Elastic Scattering | ~2 | ~0.5 | 3 | 110→100 |
| 2 | Transition | 4-8 | CID (cooling) | ~6.5 | ~1.5 | ~13 | 100→35 |
| 3 | Negative Glow | 8-13.5 | CID (Removal) + Elastic | ~2 | ~12 | ~3 | 35→47 |

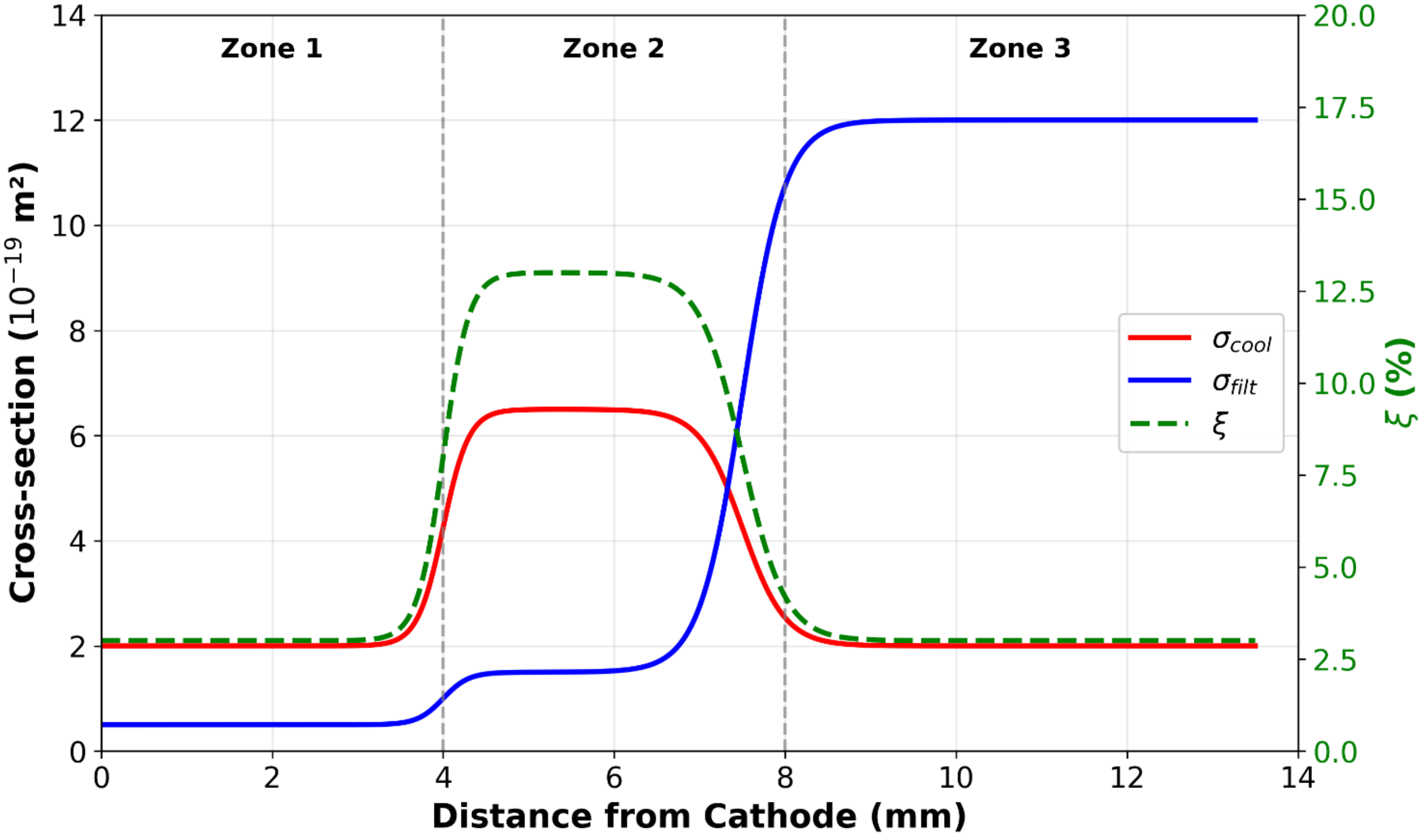

Figure 16. Spatially resolved transport parameters derived from the kinetic model. $\sigma_{cool}$ (red): cooling cross-section, enhanced in Zone 2 due to inelastic H-$CH_4$ collisions. $\sigma_{filt}$ (blue): geometric filtration cross-section, increasing dramatically in Zone 3 where velocity-selective removal

produces the observed beam hardening. ξ (green dashed, right axis): fractional energy loss per collision, elevated in Zone 2 (~13%) compared to Ar-dominated zones (~3%). The return of $\sigma_{cool}$ and ξ to Zone 1 values in Zone 3 confirms the common H-Ar collision physics in both Ar-dominated regions; the dramatically different energy evolution arises from the change in $\sigma_{filt}$.

**References**

1. Mallapragada, D. S. Decarbonization of the chemical industry through electrification: Barriers and opportunities.
2. Yuan, X. *et al.* A kinetic study of nonthermal plasma pyrolysis of methane: Insights into hydrogen and carbon material production. *Chemical Engineering Journal* **499**, 156396 (2024).
3. Heintze, M. & Magureanu, M. Methane conversion into acetylene in a microwave plasma: Optimization of the operating parameters. *J. Appl. Phys.* **92**, 2276–2283 (2002).
4. Daghagheleh, O. Optimizing methane plasma pyrolysis for instant hydrogen and high-quality carbon production. *International Journal of Hydrogen Energy* (2024).
5. Navascués, P., Cotrino, J., González-Elipe, A. R. & Gómez-Ramírez, A. Plasma assisted dry reforming of methane: Syngas and hydrocarbons formation mechanisms. *Fuel Processing Technology* **248**, 107827 (2023).
6. Heijkers, S., Aghaei, M. & Bogaerts, A. Plasma-Based CH4 Conversion into Higher Hydrocarbons and H2: Modeling to Reveal the Reaction Mechanisms of Different Plasma Sources. *J Phys Chem C Nanomater Interfaces* **124**, 7016–7030 (2020).
7. Zhou, D., McCauley, T. G., Qin, L. C., Krauss, A. R. & Gruen, D. M. Synthesis of nanocrystalline diamond thin films from an Ar–CH4 microwave plasma. *J. Appl. Phys.* **83**, (1998).
8. Szablowski, L., Wojcik, M. & Dybinski, O. Review of steam methane reforming as a method of hydrogen production. *Energy* **316**, 134540 (2025).
9. Ziółkowski, M. *et al.* Modeling the electron-impact dissociation of methane. *The Journal of Chemical Physics* **137**, 22A510 (2012).
10. Abiev, R. S., Sladkovskiy, D. A., Semikin, K. V., Murzin, D. Y. & Rebrov, E. V. Non-Thermal Plasma for Process and Energy Intensification in Dry Reforming of Methane. *Catalysts* **10**, 1358 (2020).
11. Rocca, J. J., Meyer, J. D., Farrell, M. R. & Collins, G. J. Glow-discharge-created electron beams: Cathode materials, electron gun designs, and technological applications. *J. Appl. Phys.* **56**, 790–797 (1984).
12. Kline, L. E. Pulsed Glow Discharges in Laser Excitation and Breakdown. *IEEE Trans. Elect. Insul.* **EI-17**, 121–124 (1982).
13. Winchester, M. R. & Payling, R. Radio-frequency glow discharge spectrometry:: A critical review. *Spectrochimica Acta Part B: Atomic Spectroscopy* **59**, 607–666 (2004).
14. Steiner, R. E., Barshick, C. M. & Bogaerts, A. Glow Discharge Optical Spectroscopy and Mass Spectrometry. in *Encyclopedia of Analytical Chemistry* (John Wiley & Sons, Ltd, 2009). doi:10.1002/9780470027318.a5107.pub2.
15. Wang, X., Zhou, M. & Jin, X. Application of glow discharge plasma for wastewater treatment. *Electrochimica Acta* **83**, 501–512 (2012).

16. Trenchev, G., Nikiforov, A., Wang, W., Kolev, St. & Bogaerts, A. Atmospheric pressure glow discharge for CO2 conversion: Model-based exploration of the optimum reactor configuration. *Chemical Engineering Journal* **362**, 830–841 (2019).
17. Montie, T. C., Kelly-Wintenberg, K. & Roth, J. R. An overview of research using the one atmosphere uniform glow discharge plasma (OAUGDP) for sterilization of surfaces and materials. *IEEE Transactions on Plasma Science* **28**, 41–50 (2000).
18. Hopf, C., Jacob, W. & Rohde, V. Oxygen glow discharge cleaning in nuclear fusion devices. *Journal of Nuclear Materials* **374**, 413–421 (2008).
19. Farouk, T., Farouk, B., Gutsol, A. & Fridman, A. Atmospheric pressure methane–hydrogen dc micro-glow discharge for thin film deposition. *J. Phys. D: Appl. Phys.* **41**, 175202 (2008).
20. Hartmann, P., Haubner, R. & Lux, B. Deposition of thick diamond films by pulsed d.c. glow discharge CVD. *Diamond and Related Materials* **5**, 850–856 (1996).
21. Heiman, A. *et al.* Evolution and properties of nanodiamond films deposited by direct current glow discharge. *J. Appl. Phys.* **89**, 2622–2630 (2001).
22. Linnik, S. A. & Gaydaychuk, A. V. Processes and parameters of diamond films deposition in AC glow discharge. *Diamond and Related Materials* **32**, 43–47 (2013).
23. Gudmundsson, J. T. & Hecimovic, A. Foundations of DC plasma sources. *Plasma Sources Sci. Technol.* **26**, 123001 (2017).
24. Staack, D., Farouk, B., Gutsol, A. & Fridman, A. DC normal glow discharges in atmospheric pressure atomic and molecular gases. *Plasma Sources Sci. Technol.* **17**, 025013 (2008).
25. Kudryavtsev, A. A., Morin, A. V. & Tsendin, L. D. Role of nonlocal ionization in formation of the short glow discharge. *Tech. Phys.* **53**, 1029 (2008).
26. Kolobov, V. I. & Tsendin, L. D. Analytic model of the cathode region of a short glow discharge in light gases. *Phys. Rev. A* **46**, 7837–7852 (1992).
27. Yatom, S. *et al.* Synthesis of nanoparticles in carbon arc: measurements and modeling. *MRS Communications* **8**, 842–849 (2018).
28. Vekselman, V., Feurer, M., Huang, T., Stratton, B. & Raitses, Y. Complex structure of the carbon arc discharge for synthesis of nanotubes. *Plasma Sources Sci. Technol.* **26**, 065019 (2017).
29. Yatom, S. & Raitses, Y. Characterization of plasma and gas-phase chemistry during boron-nitride nanomaterial synthesis by laser-ablation of boron-rich targets. *Physical Chemistry Chemical Physics* **22**, 20837–20850 (2020).
30. Jayanarasimhan, A. *et al.* Insights into Sustainable Nitrogen Fixation by Gas-phase Spectroscopic Measurements and Global Modeling of Reaction Intermediates in Humid Nitrogen Plasma. *ACS Sustainable Chem. Eng.* **13**, 140–150 (2025).
31. Nikhar, T. *et al.* Importance of gas heating in capacitively coupled radiofrequency plasma-assisted synthesis of carbon nanomaterials. *J. Phys. D: Appl. Phys.* **57**, 475205 (2024).
32. Niemi, K., Gathen, V. S. der & Döbele, H. F. Absolute calibration of atomic density measurements by laser-induced fluorescence spectroscopy with two-photon excitation. *J. Phys. D: Appl. Phys.* **34**, 2330–2335 (2001).

33. Yue, Y., Kondeti, V. S. S. K. & Bruggeman, P. J. Absolute atomic hydrogen density measurements in an atmospheric pressure plasma jet: generation, transport and recombination from the active discharge region to the effluent. *Plasma Sources Sci. Technol.* **29**, 04LT01 (2020).
34. Kondeti, V. S. S. K., Gangal, U., Yatom, S. & Bruggeman, P. J. Ag+ reduction and silver nanoparticle synthesis at the plasma–liquid interface by an RF driven atmospheric pressure plasma jet: Mechanisms and the effect of surfactant. *Journal of Vacuum Science & Technology A* **35**, 061302 (2017).
35. Luque, J., Juchmann, W. & Jeffries, J. B. Spatial density distributions of C2, C3, and CH radicals by laser-induced fluorescence in a diamond depositing dc-arcjet. *Journal of Applied Physics* **82**, 2072–2081 (1997).
36. Dzierżęga, K., Volz, U., Nave, G. & Griesmann, U. Accurate transition rates for the 5 p – 5 s transitions in Kr I. *Phys. Rev. A* **62**, 022505 (2000).
37. Yatom, S., Luo, Y., Xiong, Q. & Bruggeman, P. J. Nanosecond pulsed humid Ar plasma jet in air: shielding, discharge characteristics and atomic hydrogen production. *J. Phys. D: Appl. Phys.* **50**, 415204 (2017).
38. Luque, J., Juchmann, W. & Jeffries, J. B. Absolute concentration measurements of CH radicals in a diamond-depositing dc-arcjet reactor. *Appl. Opt.* **36**, 3261 (1997).
39. Luque, J. & Crosley, D. R. Absolute CH concentrations in low-pressure flames measured with laser-induced fluorescence. *Appl. Phys. B* **63**, 91–98 (1996).
40. Luque, J., Smith, G. P. & Crosley, D. R. Quantitative CH determinations in low-pressure flames. *Symposium (International) on Combustion* **26**, 959–966 (1996).
41. Luque, J., Klein-Douwel, R. J. H., Jeffries, J. B., Smith, G. P. & Crosley, D. R. Quantitative laser-induced fluorescence of CH in atmospheric pressure flames. *Applied Physics B: Lasers and Optics* **76**, 715–715 (2003).
42. Carter, C. D., Donbar, J. M. & Driscoll, J. F. Simultaneous CH planar laser-induced fluorescence and particle imaging velocimetry in turbulent nonpremixed flames. *Appl Phys B* **66**, 129–132 (1998).
43. Klein-Douwel, R. J. H., Spaanjaars, J. J. L. & ter Meulen, J. J. Two-dimensional distributions of $C_2$, CH, and OH in a diamond depositing oxyacetylene flame measured by laser induced fluorescence. *Journal of Applied Physics* **78**, 2086–2096 (1995).
44. Kaminski, C. & Ewart, P. Absolute concentration measurements of C2 in a diamond CVD reactor by laser-induced fluorescence. *Appl. Phys. B* **61**, 585–592 (1995).
45. Yatom, S. Measurement and reduction of Ar metastable densities by nitrogen admixing in electron beam-generated plasmas. *Plasma Sources Sci. Technol.* **32**, 115005 (2023).
46. Yatom, S. & Dobrynin, D. Examination of OH and H2O2 production by uniform and non-uniform modes of dielectric barrier discharge in He/air mixture. *J. Phys. D: Appl. Phys.* **55**, 485203 (2022).
47. Bauer, W., Becker, K. H., Bielefeld, M. & Meuser, R. Lifetime measurements on electronically excited C2(A1Πu) and C2(d3Πg) by laser-induced fluorescence. *Chemical Physics Letters* **123**, 33–36 (1986).

48. Naulin, C., Costes, M. & Dorthe, G. C2 Radicals in a supersonic molecular beam. Radiative lifetime of the d 3Πg state measured by laser-induced fluorescence. *Chemical Physics Letters* **143**, 496–500 (1988).
49. Colgate, S. O. & Schmidt, T. W. Energy-Dependence Measurement of the CH4++CH4=CH3+CH5+ Reaction Cross Section. *J. Chem. Phys.* **45**, 367–369 (1966).
50. Peko, B. L., Dyakov, I. V. & Champion, R. L. Collision induced dissociation, proton abstraction, and charge transfer for low energy collisions involving CH4+. *J. Chem. Phys.* **109**, 5269–5275 (1998).
51. Rapp, D. & Francis, W. E. Charge Exchange between Gaseous Ions and Atoms. *J. Chem. Phys.* **37**, 2631–2645 (1962).
52. Ganguly, B. N. & Garscadden, A. Electric field and Doppler emission profile measurements in an obstructed hydrogen discharge. *Journal of Applied Physics* **70**, 621–627 (1991).
53. Cappelli, A. L., Gottscho, R. A. & Miller, T. A. Doppler-broadened line shapes of atomic hydrogen in a parallel-plate radio frequency discharge. *Plasma Chem Plasma Process* **5**, 317–331 (1985).
54. Adamov, M. R. G., Obradovic, B. M., Kuraica, M. M. & Konjevic, N. Doppler spectroscopy of hydrogen and deuterium Balmer alpha line in an abnormal glow discharge. *IEEE Transactions on Plasma Science* **31**, 444–454 (2003).
55. Cvetanović, N., Kuraica, M. M. & Konjević, N. Excessive Balmer line broadening in a plane cathode abnormal glow discharge in hydrogen. *Journal of Applied Physics* **97**, 033302 (2005).
56. Petrović, Z. Lj., Jelenković, B. M. & Phelps, A. V. Excitation by and surface reflection of fast hydrogen atoms in low-pressure hydrogen discharges. *Phys. Rev. Lett.* **68**, 325–328 (1992).
57. Tatarova, E. *et al.* Hot and super-hot hydrogen atoms in microwave plasma. *Applied Physics Letters* **95**, 181503 (2009).
58. Spasojevic, D. & Steflekova, V. Electric field distribution in the cathode-fall region of an abnormal glow discharge in hydrogen: experiment and theory. *Plasma Sources Sci. Technol.* (2012).
59. Spasojević, Dj., Cvejić, M., Šišović, N. M. & Konjević, N. Simultaneous plasma and electric field diagnostics of microdischarge from hydrogen Balmer line shape. *Appl. Phys. Lett.* **96**, 241501 (2010).
60. Wujec, T. & Janus, H. W. Spectroscopic measurements of electric field distributions in dielectric barrier discharges in hydrogen.
61. Lieberman, M. A. & Lichtenberg, A. J. *Principles of Plasma Discharges and Materials Processing: Lieberman/Plasma 2e*. (John Wiley & Sons, Inc., Hoboken, NJ, USA, 2005). doi:10.1002/0471724254.
62. Van Surksum, T. L. & Fisher, E. R. Gas-phase diagnostic studies of H2 and CH4 inductively coupled plasmas. *J. Vac. Sci. Technol. A* **38**, 033010 (2020).
63. Luque, J. *et al.* Gas temperature measurement in CH4/CO2 dielectric-barrier discharges by optical emission spectroscopy. *Journal of Applied Physics* **93**, 4432–4438 (2003).

64. Avtaeva, S. V. & Lapochkina, T. M. Characteristics of molecular hydrogen and CH* radicals in a methane plasma in a magnetically enhanced capacitive RF discharge. *Plasma Phys. Rep.* **33**, 774–785 (2007).
65. Sadiek, I., Di Bernardo, S., Macherius, U. & Van Helden, J.-P. H. Formation of HNC and HCN isomers in molecular plasmas revealed by frequency comb and quantum cascade laser spectroscopy. *Phys. Chem. Chem. Phys.* **28**, 3929–3938 (2026).
66. Phelps, A. V. Cross Sections and Swarm Coefficients for H+, H2+, H3+, H, H2, and H− in H2 for Energies from 0.1 eV to 10 keV. *J. Phys. Chem. Ref. Data* **19**, (1990).
67. Krstić, P. S. & Schultz, D. R. Elastic and related transport cross sections for protons scattering from the noble gases He, Ne, Ar, Kr, and Xe. *Physics of Plasmas* **13**, 053501 (2006).
68. Phelps, A. V. Cross Sections and Swarm Coefficients for Nitrogen Ions and Neutrals in N2 and Argon Ions and Neutrals in Ar for Energies from 0.1 eV to 10 keV. *Journal of Physical and Chemical Reference Data* **20**, 557–573 (1991).
69. Shirai, T., Tabata, T., Tawara, H. & Itikawa, Y. ANALYTIC CROSS SECTIONS FOR ELECTRON COLLISIONS WITH HYDROCARBONS: CH4, C2H6, C2H4, C2H2, C3H8, AND C3H6. *Atomic Data and Nuclear Data Tables* **80**, 147–204 (2002).
70. Song, M.-Y. *et al.* Cross Sections for Electron Collisions with Methane. *J. Phys. Chem. Ref. Data* **44**, 023101 (2015).
71. Ruzic, D. N. & Cohen, S. A. Total scattering cross sections and interatomic potentials for neutral hydrogen and helium on some noble gases. *The Journal of Chemical Physics* **83**, 5527–5530 (1985).
72. Janev, R. K., Langer, W. D., Post, D. E. & Evans, K. *Elementary Processes in Hydrogen-Helium Plasmas*. (Springer Berlin Heidelberg, Berlin, Heidelberg, 1987). doi:10.1007/978-3-642-71935-6.
73. Chapman, W. B., Schiffman, A., Hutson, J. M. & Nesbitt, D. J. Rotationally inelastic scattering in CH4+He, Ne, and Ar: State-to-state cross sections via direct infrared laser absorption in crossed supersonic jets. *J. Chem. Phys.* **105**, 3497–3516 (1996).
74. Danko, M. *et al.* Electron impact excitation of methane: determination of appearance energies for dissociation products. *J. Phys. B: At. Mol. Opt. Phys.* **46**, 045203 (2013).
75. Sheridan, T. E. & Goree, J. Collisional plasma sheath model. *Physics of Fluids B: Plasma Physics* **3**, 2796–2804 (1991).
76. Kuraica, M. & Konjević, N. Line shapes of atomic hydrogen in a plane-cathode abnormal glow discharge. *Phys. Rev. A* **46**, 4429–4432 (1992).
77. Cvetanović, N., Ivković, S. S., Obradović, B. M. & Kuraica, M. M. Simultaneous influence of Stark effect and excessive line broadening on the Hα line. *Eur. Phys. J. D* **71**, 317 (2017).
78. *Sputtering by Particle Bombardment: Experiments and Computer Calculations from Threshold to MeV Energies*. (Springer, Berlin, Heidelberg, 2007).
79. Eckstein, W. & Heifetz, D. B. Data sets for hydrogen reflection and their use in neutral transport calculations. *Journal of Nuclear Materials* **145–147**, 332–338 (1987).
80. Šišović, N. M., Majstorović, G. Lj. & Konjević, N. Excessive hydrogen and deuterium Balmer lines broadening in a hollow cathode glow discharges. *Eur. Phys. J. D* **32**, 347–354 (2005).

81. Cvetanović, N., Obradović, B. M. & Kuraica, M. M. Monte Carlo simulation for excessive Balmer line broadening generated by transport of fast H Atoms in an abnormal glow discharge. *Journal of Applied Physics* **105**, 043306 (2009).
82. Phelps, A. V. Collisions of H, and with Ar and of and with for Energies from 0.1 eV to 10 keV. *J. Phys. Chem. Ref. Data* **21**, (1992).
83. Finn, T. G., Carnahan, B. L., Wells, W. C. & Zipf, E. C. Dissociation of CH4 and CD4 by electron impact: Production of metastable and high-Rydberg hydrogen and carbon fragments. *J. Chem. Phys.* **63**, 1596–1604 (1975).
84. Beenakker, C. I. M. & Heer, F. J. D. Dissociative excitation of some aliphatic hydrocarbons by electron impact. *Chemical Physics* **7**, 130–136 (1975).
85. Schiavone, J. A., Donohue, D. E. & Freund, R. S. Molecular dissociation by electron impact: High Rydberg fragments from methane, ethylene, and ethane. *The Journal of Chemical Physics* **67**, 759–768 (1977).
86. Motohashi, K., Soshi, H., Ukai, M. & Tsurubuchi, S. Dissociative excitation of CH4 by electron impact: Emission cross sections for the fragment species. *Chemical Physics* **213**, 369–384 (1996).
87. Baulch, D. L., Bowman, C. T., Cobos, C. J. & Cox, R. A. Evaluated Kinetic Data for Combustion Modeling: Supplement II. *J. Phys. Chem. Ref. Data* **34**, (2005).
88. Zhou, Y., Fu, B., Wang, C., Collins, M. A. & Zhang, D. H. *Ab initio* potential energy surface and quantum dynamics for the H + CH4 → H2 + CH3 reaction. *The Journal of Chemical Physics* **134**, 064323 (2011).
89. Heijkers, S., Aghaei, M. & Bogaerts, A. Plasma-Based $CH_4$ Conversion into Higher Hydrocarbons and $H_2$ : Modeling to Reveal the Reaction Mechanisms of Different Plasma Sources. *J. Phys. Chem. C* **124**, 7016–7030 (2020).
90. Kruse, T. & Roth, P. Kinetics of C2 Reactions during High-Temperature Pyrolysis of Acetylene. *J. Phys. Chem. A* **101**, 2138–2146 (1997).
91. Morton, R. J. & Kaiser, R. I. Kinetics of suprathermal hydrogen atom reactions with saturated hydrides in planetary and satellite atmospheres. *Planetary and Space Science* **51**, 365–373 (2003).
92. Marques, J. M. C., Martínez-Núñez, E. & Vázquez, S. A. Trajectory Dynamics Study of Collision-Induced Dissociation of the Ar + CH4 Reaction at Hyperthermal Conditions: Vibrational Excitation and Isotope Substitution. *J. Phys. Chem. A* **110**, 7113–7121 (2006).
93. Bechtel, H. A., Kim, Z. H., Camden, J. P. & Zare, R. N. Bond and mode selectivity in the reaction of atomic chlorine with vibrationally excited CH2D2. *J. Chem. Phys.* **120**, 791–799 (2004).
94. Yan, S., Wu, Y.-T., Zhang, B., Yue, X.-F. & Liu, K. Do Vibrational Excitations of $CHD_3$ Preferentially Promote Reactivity Toward the Chlorine Atom? *Science* **316**, 1723–1726 (2007).
95. Crim, F. F. Vibrational State Control of Bimolecular Reactions: Discovering and Directing the Chemistry. *Acc. Chem. Res.* **32**, 877–884 (1999).
96. Bruggeman, P. J., Sadeghi, N., Schram, D. C. & Linss, V. Gas temperature determination from rotational lines in non-equilibrium plasmas: a review. *Plasma Sources Sci. Technol.* **23**, 023001 (2014).

97. Godyak, V. A. Electron energy distribution function control in gas discharge plasmas. *Phys. Plasmas* **20**, 101611 (2013).
98. Carman, R. J. Influence of the energy spectrum of electrons ejected from the cathode on electron behaviour in the cathode sheath region of a DC glow discharge in helium and in argon. *J. Phys. D: Appl. Phys.* **25**, 198 (1992).
99. Su, T. & Chesnavich, W. J. Parametrization of the ion–polar molecule collision rate constant by trajectory calculations. *J. Chem. Phys.* **76**, 5183–5185 (1982).
100. Anicich, V. An Index of the Literature for Bimolecular Gas Phase Cation-Molecule Reaction Kinetics. *JPL Publication* **03–19**, (2003).
101. Gioumousis, G. & Stevenson, D. P. Reactions of Gaseous Molecule Ions with Gaseous Molecules. V. Theory. *J. Chem. Phys.* **29**, 294–299 (1958).
102. Simón-Manso, Y. Ion-Neutral Collision Cross Section as a Function of the Static Dipole Polarizability and the Ionization Energy of the Ion. *J. Phys. Chem. A* **127**, 3274–3280 (2023).
103. Anicich, V. G. Evaluated Bimolecular Ion-Molecule Gas Phase Kinetics of Positive Ions for Use in Modeling Planetary Atmospheres, Cometary Comae, and Interstellar Clouds. *J. Phys. Chem. Ref. Data* **22**, 1469–1569 (1993).
104. Errea, L., Macías, A., Méndez, L., Rabadán, I. & Riera, A. Anisotropy effects in H++H2 collisions. *IJMS* **3**, 142–161 (2002).